\documentclass[bibyear]{aa}  

\usepackage{graphicx}
\usepackage{natbib}
\bibpunct{(}{)}{;}{a}{}{,} 
\usepackage{txfonts}
\usepackage{caption}
\begin{document}

   \title{Galactic Dark Matter:\\ a Dynamical Consequence of Cosmological Expansion}


   \author{Elias A. S. M\'egier
	}

   \institute{
              \email{elias.megier@gmail.com}
             }


 
  \abstract
  {}
   {This work wants to show how standard General Relativity is able to explain galactic rotation curves without the need for dark matter, this starting from the idea that when Einstein's equations are applied to the dynamics of a galaxy embedded in an expanding universe they do not reduce to Poisson's equation but a generalisation of it taking cosmological expansion into account.}
   {A non-linear scheme to perturb Einstein's field equations around the Robertson-Walker metric is devised in order to find their non-relativistic limit without losing their characteristic non-linearities. The resulting equation is used to numerically study the gravitational potential of a cosmological perturbation and applied to a simple galactic model with an exponentially decreasing baryonic matter distribution.}
   {The non-relativistic limit of General Relativity in a Robertson-Walker space-time produces a generalised Poisson equation for the gravitational potential which is non-linear, parabolic and heat-like. It is shown how its non-linearities generate an effective "dark matter" distribution caused by both cosmological expansion and the dynamics of the perturbation's gravitational potential. It is also shown how this dynamical effect gets completely lost during a linearisation of Einstein's equations. The equation is then used to successfully fit real galactic rotation curves numerically using a matter distribution following the shape of a simple S\'ersic luminosity profile, common to most galaxies, thus without recourse to dark matter. A relation for the dark to luminous matter ratio is found, explaining the domination of dark matter in low-mass galaxies.  A few rotation curves with a faster than Newtonian decrease are also presented and successfully fitted, opening the way to a new possible interpretation of these phenomena in terms of an effective "anti-gravitational" dark matter distribution, purely geometrical in origin.}
   {}


   \maketitle
%

\begin{flushright}{\slshape 
From my point of view one cannot arrive, by way of theory, at any at least somewhat reliable results in the field of cosmology, if one makes no use of the principle of relativity.} \\ \medskip
---Albert Einstein \citep{Ein}
\end{flushright}

\section{Introduction}

One amongst the greatest puzzles of modern astrophysics is the general asymptotic flatness of galactic rotation curves. These display a significant deviation from Newtonian theoretical curves computed using the observationally inferred matter distribution, for, according to Newtonian gravitation, their asymptotic behaviour should be that of a $r^{-1/2}$ decrease. Because of the wide belief that the General theory of Relativity should add no new element to this prediction, the flatness of rotation curves, along with other phenomena such as gravitational lensing effects, see e.g. \citep{Lense}, has led to think that, if General Relativity is to stand as a valid gravitational theory in the field of cosmology, there must be some kind of exotic and as yet unseen "dark matter" forming "dark halos" around galaxies, this in order to adjust theoretical predictions to observational facts, see e.g. \cite{Bert} for a thorough review. An other approach to the problem has also been to modify General Relativity in such a way as to accommodate observations, see e.g. \citet{Milgr,Beck}.

In contrast to this approach, I consider here a point that seems to have been neglected in the past relativistic analysis of the problem, that is, that the Universe's expansion might play a role of foremost importance in the non-relativistic approximation of General Relativity. While in a nearly Minkowskian space-time Einstein's equations reduce to Poisson's equation, in a nearly Robertson-Walker space-time they are found to reduce to a parabolic heat-like equation, displaying thus a time derivative which is completely absent in the nearly Minkowskian case. The solutions of this equation are thence time dependent, like the space-time metric itself, and can sensibly depart from their Minkowskian counterparts in their qualitative behaviour.  

An important starting idea is that many interesting features of gravitation arise from the high non-linearity of Einstein's field equations. If one is then to perturb the Robertson-Walker metric in order to study the behaviour of density fluctuations above the cosmological mean, such as galaxies, one suspects that some of the key features might get lost in a linearisation approach. 

One then follows the simple strategy of studying the full Einstein equations for the time-time component $g_{00}=-V^2$ of the Robertson-Walker metric tensor neglecting all terms quadratic in Hubble's "constant" $H$ and in the gravitational potential.\footnote{Throughout this work $H$ is treated as a constant, one must however not forget that in equation \eqref{key} it is a function of cosmological time. }  This programme leads, as we shall see, to the following modified Poisson equation:
\begin{equation}
\frac{3H}{c^2}\frac{\partial_t V}{V^3}+\frac{\nabla^2 V}{V} = \frac{4\pi G}{c^2} \rho
\label{key}
\end{equation}
where $\nabla^2$ is the Laplacian on a Robertson-Walker space-time.

Albeit the term proportional to $H$ might seem irrelevant after a na\"ive dimensional analysis, it is shown how in fact it can be \emph{dominating}, acting thus as an effective dark matter density
\begin{equation}
\rho_D=-\frac{3H}{4\pi G}\frac{\partial_t V}{V^3}\;\;.
\end{equation}
Let us note at once how this dynamical explanation of dark matter-related phenomena can take into account a wider range of cases than the dark matter conjecture, namely cases in which $\rho_D$ is \emph{negative}. Such cases might exist as we shall later see.

In what follows it will be shown how this equation can explain the flatness of rotation curves and easily fit experimental data using an exponentially decreasing matter distribution. Exponential profiles are the simplest density distributions fitting a wide range of galactic luminosity profiles \citep{Sers} and are thus the best candidates for a first dark matter free analysis.

It will also be shown how the predictions of this model, had one linearised equation \eqref{key}, would not have significantly departed from Newtonian theory. This makes clear how the non-linearities of General Relativity are crucial in a perturbation study.

An other point treated regards the fact that low-mass galaxies are dominated by their "dark mass", see e.g. \cite{Fornax, Luminosity, Burk}, this model explains this phenomenon quite naturally thanks to an expression for $\frac{\rho_D}{\rho}$ hereby exposed. This expression also explains the relation presented by \cite{Luminosity} telling that $\frac{\rho_D}{\rho}$ scales inversely with galactic luminosity.

A further point treated will be that of the so called "core/cusp controversy". Cold dark matter models predict that dark matter profiles should have a universal $r^{-a}$ cusp behaviour at the centre of galaxies with $a=1$, see e.g. \citep{CDM}, whilst in low-mass spiral galaxies many authors find a flat $a=0$ behaviour, see again \cite{Luminosity,Fornax} amongst others. In this work results agree with the latter opinion, which sheds an alternative light on the problem.

\section{The perturbation method}

The key feature one wants to add to the Robertson-Walker geometry is the presence of local acceleration fields giving rise to galactic formation and dynamics. The simplest possible degree of freedom one can introduce is then a Newtonian potential $\Phi$ such that a freely falling body with non-relativistic $4$-velocity $\mathbf{u}$ with respect to local matter may express his geodesic equation approximately as:
\begin{equation}
		\frac{d}{dt}u^i=-\partial^i\Phi\qquad\text{and}\qquad\frac{d}{dt}u^0=0
\label{geod}
\end{equation}
using some orthonormal reference frame (tetrad) $\big\lbrace\mathbf{e}_{\mu}\big\rbrace_{\mu=0,\cdots,3}$ in which to write $\mathbf{u}=\mathbf{e}_{\mu}u^{\mu}$. Such a frame has a simple geometric meaning, the vector field $\mathbf{e}_0$ can be seen at each point in space-time as the $4$-velocity of an observer suffering an acceleration $\partial^i\Phi$ causing him to resist free fall, exactly as an observer on the surface of he Earth is prevented from falling towards the Earth's core.

After the introduction of this free parameter, the Robertson-Walker metric can be written using stereographic projection coordinates as:\footnote{Boldface characters always denote tensor objects in this work, opposed to scalars and tensor components.}
\begin{equation}
	\mathbf{g}=\eta_{\mu\nu}\mathbf{e}^{\mu}\otimes\mathbf{e}^{\nu}
	\nonumber
\end{equation}
where $\eta_{\mu\nu}=diag(-1,1,1,1)$ is the Minkowsky metric and

\begin{align}
& \mathbf{e}^{0}=V\mathbf{d} t\stackrel{\text{def}}{=}\mathrm{e}^{\Phi}\mathbf{d} t \nonumber \\
 &\mathbf{e}^{i}=a(t)\frac{\mathbf{d}x^{i}}{\bigg(1+\frac{1}{4}\kappa\delta_{jl}x^{j}x^{l}\bigg)}\stackrel{\text{def}}{=}a(t)\frac{\mathbf{d}x^{i}}{\bigg(1+\frac{1}{4}\kappa r^{2}\bigg)}
\label{basis}
\end{align}
where $a(t)$ is the cosmological expansion factor, $\kappa$ the spacial intrinsic mean curvature and $\Phi\ll1$.

This might seem a simplistic approach, it has however the virtue of making equations relatively tractable and of introducing only those elements that are truly vital in the model. 

Let us note that the case in which $a(t)=1$ and $\kappa=0$ reduces to a space-time displaying only Newtonian phenomena such as a conservative acceleration field and a deflection of light corresponding to $1/2$ of the correct Schwarzschild value. In other words, this model is expected to give correct insight only into non (special) relativistic phenomena such as galactic evolution in a Robertson-Walker space-time, exactly as its Minkowskian counterpart, which is only good at recovering ordinary Newtonian dynamics.

In this basis the connection $1$-forms read:
\[
\boldsymbol{\omega}^{0}_{\phantom{0}i}=\frac{\mathrm{e}_{i}\big(V\big)}{V}\mathbf{e}^{0}+\frac{\stackrel{\cdot}{a}(t)}{a(t)}\frac{\mathbf{e}^{i}}{V}=\frac{\mathrm{e}_{i}\big(V\big)}{V}\mathbf{e}^{0}+\frac{H}{V}\mathbf{e}^{i}
\] 
\[
\boldsymbol{\omega}^{i}_{\phantom{i}j}=\frac{k}{2a(t)}\left[x_{i}\mathbf{e}^{j}-x_{j}\mathbf{e}^{i}\right]
\] 
where $\mathrm{e}_{i}$ is the action of $\mathbf{e}_i$ as a derivative, that is $\mathrm{e}_{i}\stackrel{\text{def}}{=}\frac{\left(1+\frac{1}{4}\kappa r^2\right)}{a(t)}\partial_i$,  and the dot represents derivation with respect to t. One already sees the appearance of the term coupling $H$ and $V$ that is going to be so important in what follows.

As mentioned before, for a non relativistic particle whose $4$-velocity $\mathbf{u}\approx\mathbf{e}_0$ equation \eqref{geod} reads:
\[
	\frac{d}{dt}u^i\approx\omega^0_{\phantom{0}0i}=-\mathrm{e}_i\big(\Phi\big)=-\frac{\left(1+\frac{1}{4}\kappa r^2\right)}{a(t)}\partial_i\Phi\qquad\text{and}\qquad\frac{d}{dt}u^0\approx 0\;\;.
\]

From these connection $1$-forms one then computes the Einstein tensor $\mathbf{G}$. 

Having introduced only one degree of freedom $\Phi$ one needs one single equation to fix it in a physical way. It is time to remember that we are interested in non relativistic phenomena, in other words, Newtonian dynamics embedded in an expanding universe. This is why one assumes that the main contribution to the perturbation stress energy tensor $\mathbf{T}_p$ be the matter-energy density $\rho_p$ and may neglect its pressure and stresses. Hence, once the computed Einstein tensor has been easily and uniquely separated in two parts\footnote{Thanks to the choice of an orthonormal basis, in which tensor components acquire a direct physical meaning.}  $\mathbf{G}=\mathbf{G}_{RW}+\mathbf{G}_p$, $\mathbf{G}_{RW}$ being the Robertson-Walker Einstein tensor and $\mathbf{G}_p$ that of the perturbation, we make the correspondence $Tr\big(\mathbf{T}_p\big)\approx-\rho_p$.\footnote{The sign is due to the choice of signature for the metric.} This leads to the equation
\[
Tr\big(\mathbf{G}_p\big)=-8\pi G \rho_p
\]
which finally takes us to the relation:
\begin{align*}
& 8\pi G\rho_p  = 2 e^{-\Phi} \nabla^{2} e^{\Phi} +
6 \frac{\stackrel{\cdot}{a}(t)}{a(t)} e^{-2\Phi}\partial _t\Phi + \\
& \frac{5\kappa x^i\partial _i\Phi }{a^2(t)}\left(1+\frac{1}{4}\kappa r^2\right)
+6\left(\frac{\stackrel{\cdot}{a}(t)}{a(t)}\right)^2\left[e^{-2\Phi}-1\right]
 +6\frac{\stackrel{\cdot\cdot}{a}(t)}{a(t)}\left[ e^{-\Phi}-1\right] \\
& \phantom{8\Pi G \rho_p}=
2 e^{-\Phi} \nabla^{2} e^{\Phi} +
6 H e^{-2\Phi}\partial _t\Phi + \\
& \frac{5\kappa x^i\partial _i\Phi }{a^2(t)}\left(1+\frac{1}{4}\kappa r^2\right)
+6H^2\left[e^{-2\Phi}-1\right]
 -6qH^2\left[ e^{-\Phi}-1\right]
\end{align*}
where $q$ is the deceleration parameter.
The perturbation $\rho_p$ is taken to have a linear dimension  $D\ll|\kappa|^{-1/2}$, that is, small compared to the radius of curvature of space. Neglecting terms of order $H^2$ and higher one finds equation \eqref{key} \footnote{Only after returning to international system units.}.

It is important to note that the presence of non-linear coupling terms of the form $H\partial _t V/V^3$ is completely general for dimensional reasons and does not depend on this particular perturbation approach. Of course when allowing other degrees of freedom in the metric one will find new terms that acquire some importance when one tries to study relativistic phenomena such as gravitational lensing or when trying to enlarge the validity domain of the model.


\section{The equation}

\begin{equation*}
\frac{3H}{c^2}\frac{\partial_t V}{V^3}+\frac{\nabla^2 V}{V} = \frac{4\pi G}{c^2} \rho
\end{equation*}
We remember that this equation is valid as long as $V\approx1$. This is a non-linear parabolic heat-like partial differential equation with negative diffusivity, expressing the fact that gravitation has a concentrating action. This equation seems difficult to discuss in general.

An important feature of this equation, being heat-like, is that the interaction's propagation velocity is expected to be infinite. This is exactly what one seeks in a non-relativistic equation.

One would be tempted to use the $V\approx1$ condition to remove non-linearities, this is equivalent to a direct linearisation of Einstein's equations. This approach leads however to rotation curves with a Newtonian behaviour. This is an important point and it is, I think, the main reason that has led to the wide belief that General Relativity would not produce predictions departing significantly from Newtonian theory. It is therefore interesting to show this point explicitly with the following example.

\paragraph*{Newtonian behaviour after linearisation}
A generally tractable linearised version of \eqref{key} is
\begin{equation}
\frac{3H}{c^2}\partial_t V = \left[-\nabla^2+\frac{4\pi G}{c^2} \rho\right] V
\end{equation} which is a Schr\"odinger-like equation. It has been obtained from \eqref{key} multiplying it by $V$ and setting all  $V$ terms in the denominators to $1$ as the condition $V\approx1$ would suggest. Choosing a spherically symmetric and fast decreasing matter distribution, setting $a(t)=1$ and writing the Laplacian in spherical coordinates one has:
\begin{equation*}
\frac{3H}{c^2}\partial_t V =\left[-\frac{1}{r}\frac{d^2}{dr^2}r+\frac{4\pi G}{c^2}  \rho\right] V
\end{equation*}
defining $U\stackrel{\text{def}}{=}rV$ one finds:
\begin{equation*}
\frac{3H}{c^2}\partial_t U =\left[-\frac{d^2}{dr^2}+\frac{4\pi G}{c^2}  \rho\right] U\;\;.
\end{equation*}
This can thus be reduced to an eigenvalue equation
\begin{equation}
\left[-\frac{d^2}{dr^2}+\left(\frac{4\pi G}{c^2} \rho-E\right)\right] U_E=0
\label{lin}
\end{equation}
and one takes a solution written as\footnote{When $E>0$ the conclusions are very similar but the derivation more lengthy due to convergence issues in \eqref{super} caused by the exponential factor. The main idea to treat \eqref{super} in general is to introduce a cut-off for positive values of $E$. This does not eliminate the possible divergence as $t\rightarrow\infty$ but makes clear that the behaviour of $V$ as a function of $r$ does not make room for flat rotation curves.}
\begin{equation}
U(r,t)=\int\limits_{-\infty}^{0}\!\!\text{d}EU_E(r)\mathrm{e}^{E\frac{tc^2}{3H}}\;\;.
\label{super}
\end{equation}  
Let us introduce a new time parameter $\tau=\frac{tc^2}{3H}$. In the region where $\rho$ is negligible the general solution of \eqref{lin}  for $E<0$ reads:
\begin{equation*}
U_E=A\left(\sqrt{|E|}\right)\cos\left(r\sqrt{|E|}+\phi\left(\sqrt{|E|}\right)\right)
\end{equation*}
where $A\left(\sqrt{|E|}\right)$ and $\phi\left(\sqrt{|E|}\right)$ are arbitrary functions of $\sqrt{|E|}$. Changing integration variable in \eqref{super} to $y=\sqrt{|E|}$ one has
\begin{equation*}
U(r,\tau)=-\frac{1}{\tau}\int\limits_{0}^{\infty}\text{d}\left(\mathrm{e}^{-y^2\tau}\right)A(y)\cos\left(yr+\phi(y)\right)+C_1+C_2r
\end{equation*}for some constants $C_1$ and $C_2$ giving the usual $\tau$-independent gravitational potential.
Integrating by parts one finds
\begin{align*}
&U(r,\tau)=
-\frac{A(0)\cos\left(\phi(0)\right)}{\tau}+\\
& +\frac{1}{\tau}
\int\limits_{0}^{\infty}
\text{d}y\left\lbrace\vphantom{\frac{V}{V}}\:
 \mathrm{e}^{-y^2\tau} \left[A'(y)\cos\left(yr+\phi(y)\right)+\right.\right.\\
& \left. \left. -[r+\phi'(y)]A(y)\sin\left(yr+\phi(y)\right)\right]
\vphantom{\frac{V}{V}}\right\rbrace+C_1+C_2r\;\;.
\end{align*}
It is possible to study the asymptotic behaviour of this expression for $\tau$ sufficiently great. Let us note that in order to have an effective dark matter density of order $1M_\odot/pc^3$, as in galaxies, the interesting order of magnitude for $E$ is $10^{-45}m^{-2}$, so that $E\tau\gg1$ after a few thousand years in the region where $A$ is relevant.  Let $\nolinebreak{I(\tau)=1/\tau\int\limits_0^{\infty}\text{d}y\:\mathrm{e}^{-y^2\tau}f(y)}$ one then has:
\begin{align*}
\lim\limits_{\tau\rightarrow\infty}\tau^{3/2}I(\tau)=f(0)\sqrt{\frac{\pi}{2}}
\end{align*}
for $\lim\limits_{\tau\rightarrow\infty}\sqrt{\tau}e^{-y^2\tau}=\sqrt{\frac{\pi}{2}}\delta\left(y\right)$ in the sense of distributions.
This means that $I(\tau)$ is asymptotic to $f(0)\sqrt{\frac{\pi}{2\tau^3}}$. As a result
\begin{equation*}
U(r,\tau)\sim\frac{a}{\tau}+\frac{b+rc}{\tau^{3/2}}+C_1+C_2r=r\left[C_2 +\frac{c}{\tau^{3/2}}\right]+\bigg[C_1+\frac{a}{\tau}+\frac{b}{\tau^{3/2}}\bigg]
\end{equation*}where $a$, $b$ and $c$ are three constants derived from the profile $A(y)$. Let us note that only two of them are independent because $c/a=-\tan(\phi(0))$.

From this we see that in general
\begin{equation*}
V(r,t)=\frac{C_1(t)}{r}+C_2(t)
\end{equation*}after little time compared to galactic and cosmological evolution. This has exactly the form of a Newtonian potential and is hence incapable to produce flat rotation curves.

\paragraph*{A solution approach}

Let us now study in contrast the non-linear case. This will give completely different results thanks to the appearance of a term proportional to $r^2$ acting as an effective matter distribution that becomes important for large radii and low-mass galaxies, exactly as dark matter does. Such a term appears thanks to the non-linearities in \eqref{key} and is responsible for the qualitative change of the solutions' behaviour. This term gets completely lost during linearisation.

In what follows we assume the system being studied to be spherically symmetric for the sake of simplicity although equation \eqref{key} can be applied to any matter distribution. This simplification, however strong it may seem, influences the physical results of this work only when it comes to give precise values to galactic masses after fitting procedures. The matter distributions considered are taken to be time independent as well as all cosmological parameters involved in \eqref{key}. All results in this work are then valid in a region of space-time in which $H$ and $a$ can be approximated to constants and $V\approx1$. These conditions are fortunately soft enough to study the evolution of galactic rotation curves over cosmologically relevant periods and distances.

A fruitful approach to study \eqref{key} is to find special solutions of the form $V(r,t)=s(r)u(t)$. Substituting in \eqref{key} one finds:
\begin{equation*}
\frac{3H}{c^2}\frac{\stackrel{\cdot}{u}(t)}{u^3(t)s^2(r)}+\frac{1}{rs(r)}\left(\frac{d}{dr}\right)^2rs(r) = \frac{4\pi G}{c^2} \rho(r)
\end{equation*}
that is
\begin{equation*}
\frac{3H}{c^2}\frac{\stackrel{\cdot}{u}(t)}{u^3(t)}+\frac{s(r)}{r}\left(\frac{d}{dr}\right)^2rs(r) - \frac{4\pi G}{c^2} s^2(r)\rho(r) =0\;\;.
\end{equation*}
So, being $r$ and $t$ independent the two terms above must both be separately constant. Introducing $w(r)=rs(r)$ this becomes:
\begin{align*}
\left\lbrace
\begin{array}{c}
\stackrel{\cdot}{u}(t)=-\omega u^3(t)\phantom{w''(r)\rho(r)+\omega \frac{3H}{c^2}r^2} \\
w(r)w''(r)=\frac{4\pi G}{c^2}w^2(r)\rho(r)+\omega \frac{3H}{c^2}r^2
\end{array}\right.
\end{align*}for some frequency $\omega\in\mathbb{R}$.
From this expression one clearly sees how the above mentioned effective diverging matter distribution appears.

The equation for $u(t)$ is readily solved, one finds:
\begin{equation}
u(t)=\frac{1}{\sqrt{2\omega t+C}}
\end{equation} 
where $C$ is an integration constant that will be used later to fix $V\approx1$.

The equation for $w(r)$ is much more difficult to study in all generality and in what follows I proceed with a numerical study. But first is is interesting to study it in the region where $\rho(r)=0$  for qualitative reasons. In this region it becomes an Emden-Fowler equation and its known special solution is:
\begin{equation*}
w(r)=\sqrt{\frac{\omega}{6H}}cr^2\qquad\text{so}\qquad s(r)=\sqrt{\frac{\omega}{6H}}cr\;\;.
\end{equation*}
Computing the acceleration that the potential $\Phi=\log\left(V(r,t)\right)$ produces, one finds that a body orbiting with uniform circular motion around the origin would have a constant velocity $\mathit{v}(r)=\frac{c}{\sqrt{2}}$. Although this value is enormous we clearly see that this is the right way to follow. No Newtonian tractation could have given a \emph{constant} rotation curve at large radii.

To study the problem further one has to choose the constant matter distribution to work with. The simplest choice is to use a S\'ersic profile, for these profiles are found to fit most galactic luminosity distributions \citep{Sers,bisSers}. We assume here the total matter density to follow that of luminous matter and hence write
\begin{equation*}
\rho(r)=\rho_0 \mathrm{e}^{-\left(\frac{r}{r_0}\right)^{\frac{1}{n}}}\;\;.
\end{equation*}where $\rho_0$ is the central matter density of the galaxy, $r_0$ its characteristic radius and $n$ the S\'ersic index.

In order to proceed with numerical integration it is convenient to rewrite the equation for $w(r)$ in terms of the dimensionless variable $x=r/r_0$. This leads to
\begin{equation}
\left(\frac{d}{dx}\right)^2w(x)=\frac{4\pi G\rho_0 r_0^2}{c^2} e^{-x^{\frac{1}{n}}}w(x)+\omega \frac{3H r_0^4 }{c^2}\frac{x^2}{w(x)}\;\;.
\end{equation}The value for Hubble's constant used here is $72\:\text{km}\:\text{Mpc}^{-1}\text{s}^{-1}$.
One has to set initial value data, here the values of $w(x)$  and the value of the rotation velocity $\mathit{v}(r)$ both in $x=1$ are chosen. 

Before going any further let us write the expression for the $\frac{\rho_D}{\rho_0}$ ratio:
\begin{equation}
\frac{\rho_D}{\rho_0}=\frac{3H}{4\pi G}\frac{\omega r_0^2x^2}{\rho_0 w^2(x)}\:.
\end{equation}
As $\rho_D$ is independent of $\rho_0$ and as we have assumed $\rho$ to be proportional to the luminous matter distribution for simplicity, one clearly sees that  $\frac{\rho_D}{\rho_0}$ grows as $\rho_0$ decreases  explaining within this simple model the domination of dark matter in galaxies with a low luminous matter content, i.e. low-mass galaxies.

This formula also makes evident why in their study of a broad sample of spiral galaxies \citet{Luminosity} find $\frac{\rho_D}{\rho_0}\propto L^{-1}$ where $L$ is the luminosity of the galaxy, as it is assumed here that $\rho \propto L$.

A numerical study of this equation shows that for typical values of galactic parameters good rotation curves are obtained for extremely small values of $\omega$, indicating thus characteristic evolution times for $V(r,t)$ incomparably greater than the time scale over which \eqref{key} is expected to be valid. This reduces the role of $u(t)$ to a mere normalisation factor.

Here follow the plots of a typical rotation curve along with its corresponding potential and generating matter distribution.

\begin{center}
\includegraphics[width=200 pt]{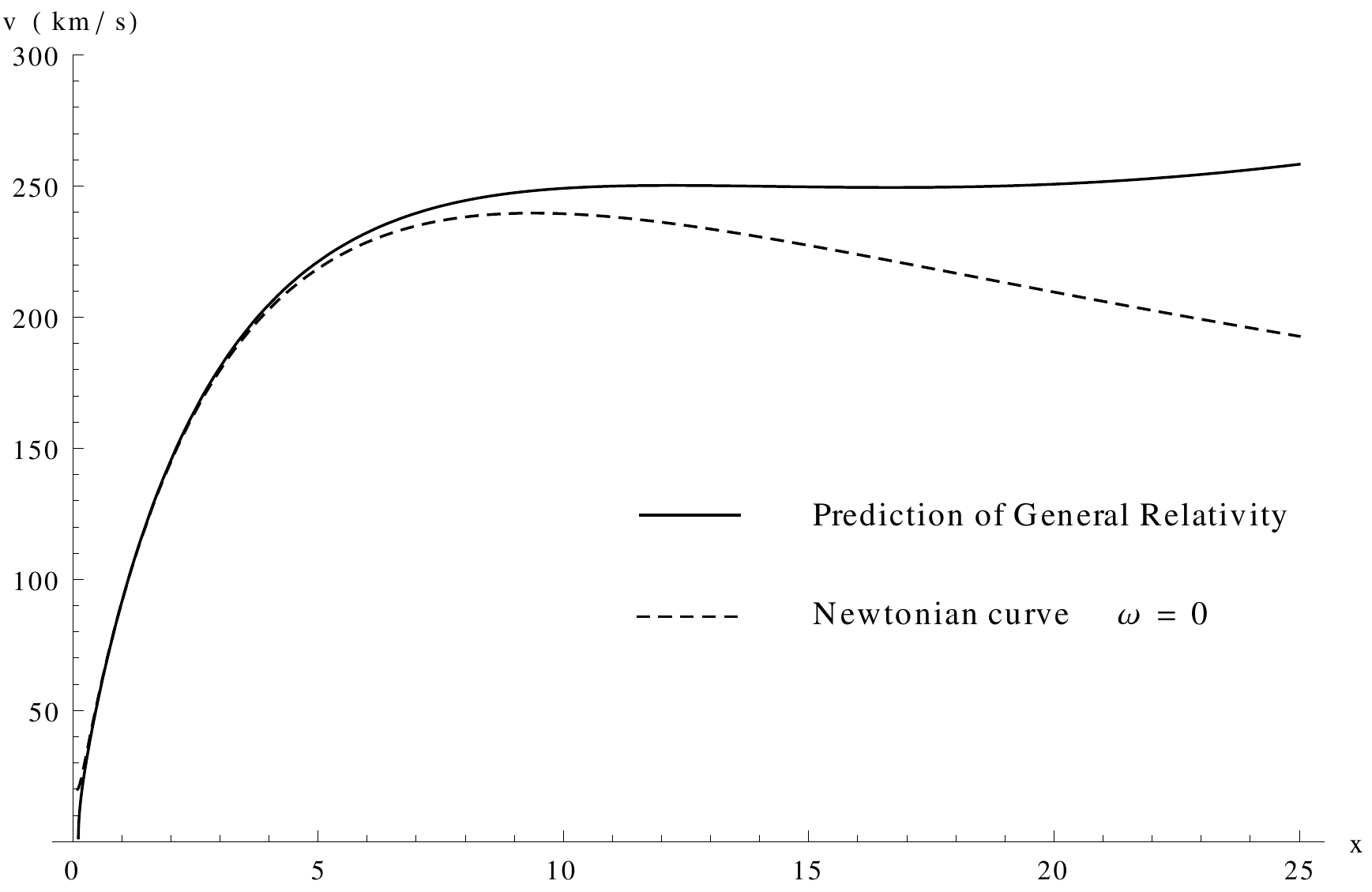}
\captionof{figure}{Typical rotation curve}\label{typical}
\pagebreak

\includegraphics[width=220 pt]{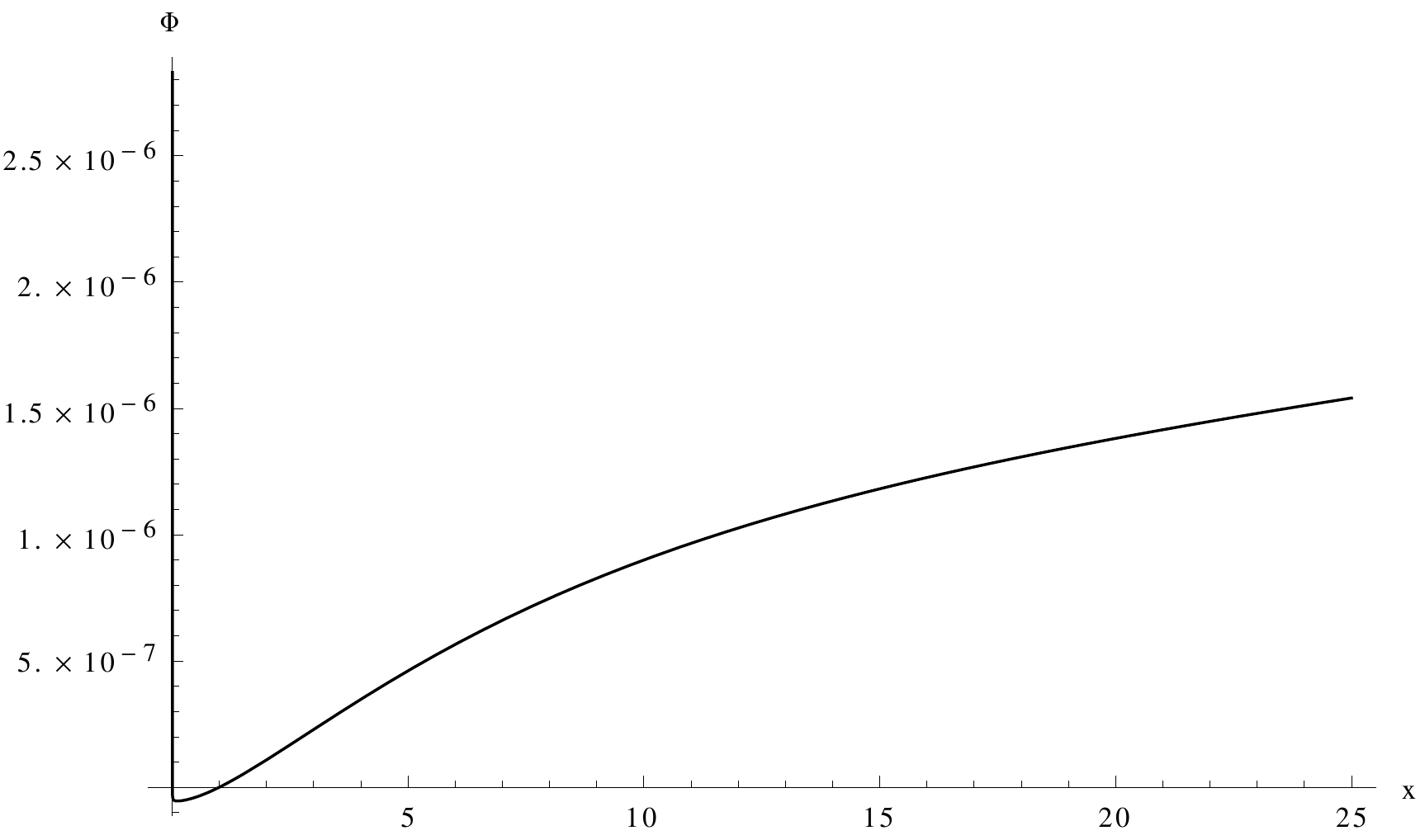} 
\captionof{figure}{Typical potential ($\Phi\approx\mathrm{e}^{\Phi}-1$)}

\nopagebreak{
\includegraphics[width=200 pt]{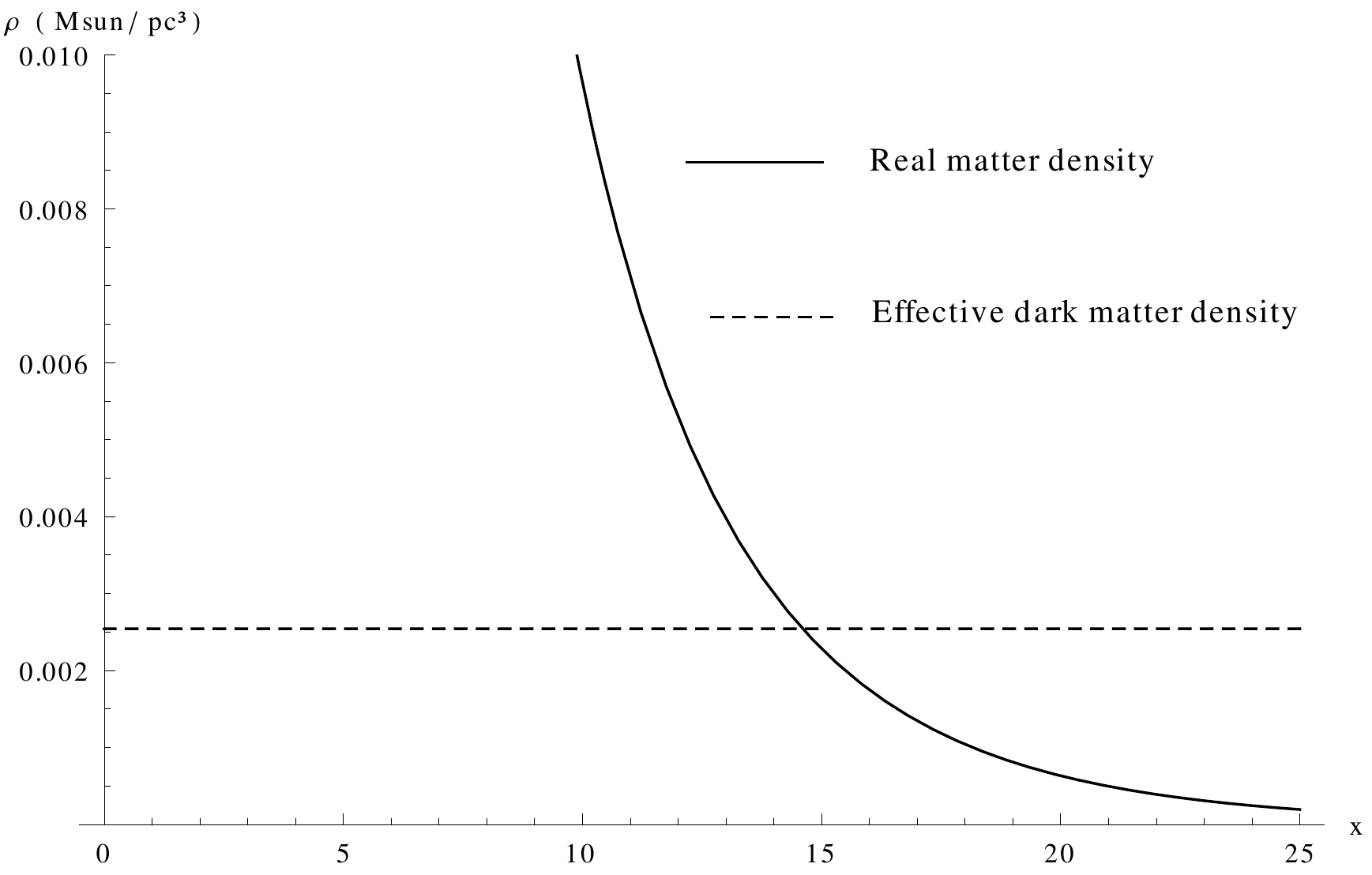} 
\captionof{figure}{Matter density}

\captionof{table}{Curve parameters}
\begin{tabular}{ l l | r l }
  $\omega$ & $2\cdot10^{-50}$ Hz & $w(1)$ & $29$ m \\
  $r_0$ & $1$ kpc & $\mathit{v}(1)$ & $91$ $kms^{-1}$ \\ \cline{3-4}
  $\rho_0$ & $1$ $M_\odot pc^{-3}$ & Total galactic mass\\
  $n$ & $3/2$ &  $2\cdot 10^{11}$ $M_\odot$  \\
\end{tabular}
}
\end{center}

 One clearly sees how the flatness of rotation curves could be explained using General Relativity in terms of a constant effective dark matter density. The plot shows a model galaxy resembling the Milky Way for which the rotation velocity keeps constant to large radii around a value of $250$ km$\text{s}^{-1}$ --- the rotation speed of the sun is about $220$ km$\text{s}^{-1}$ --- and the visible matter density halfway from the centre is of order $10^{-2}$ $M_\odot/pc^3$ very much like in the solar neighbourhood. One also sees from this model how the effective dark matter distribution becomes dominant giving a divergent total "dark mass". The behaviour of $\Phi$ shows that the condition $V=e^\Phi\approx1$ remains accurately valid throughout the region under study, assuring a posteriori the applicability of \eqref{key} to the study of galaxies.  

As mentioned in the introduction, this numerical study seems to have also a word to say regarding the so so called "core/cusp controversy". Many studies point out that in low-mass galaxies dark matter distributions near the galactic centre are fitted well by flat density profiles, exactly as in the fits presented in this work. This result disagrees with the cold dark matter paradigm as this predicts a universal $r^{-1}$ cusp behaviour near galactic centres, again see \citet{CDM} for reference.

As low-mass galaxies are completely dark matter dominated, there seem to be fewer complications in the analysis due to baryonic interactions. These galaxies are thus expected to exhibit in a clearer way purely dark matter related phenomena such as those presented in this work. The fact that in these very cases experimental observations agree with the theoretical predictions of this work is of great interest. 

This model is hence fully able to explain flat rotation curves, let us now see how it opens the way to other interesting phenomena such as "anti-gravitational dark matter". This is nothing but a negative effective dark matter density to be found when $\omega<0$. This causes a decrease of rotation curves which is faster than Newtonian, as illustrated in the following plots. 
\begin{center}

\includegraphics[width=200 pt]{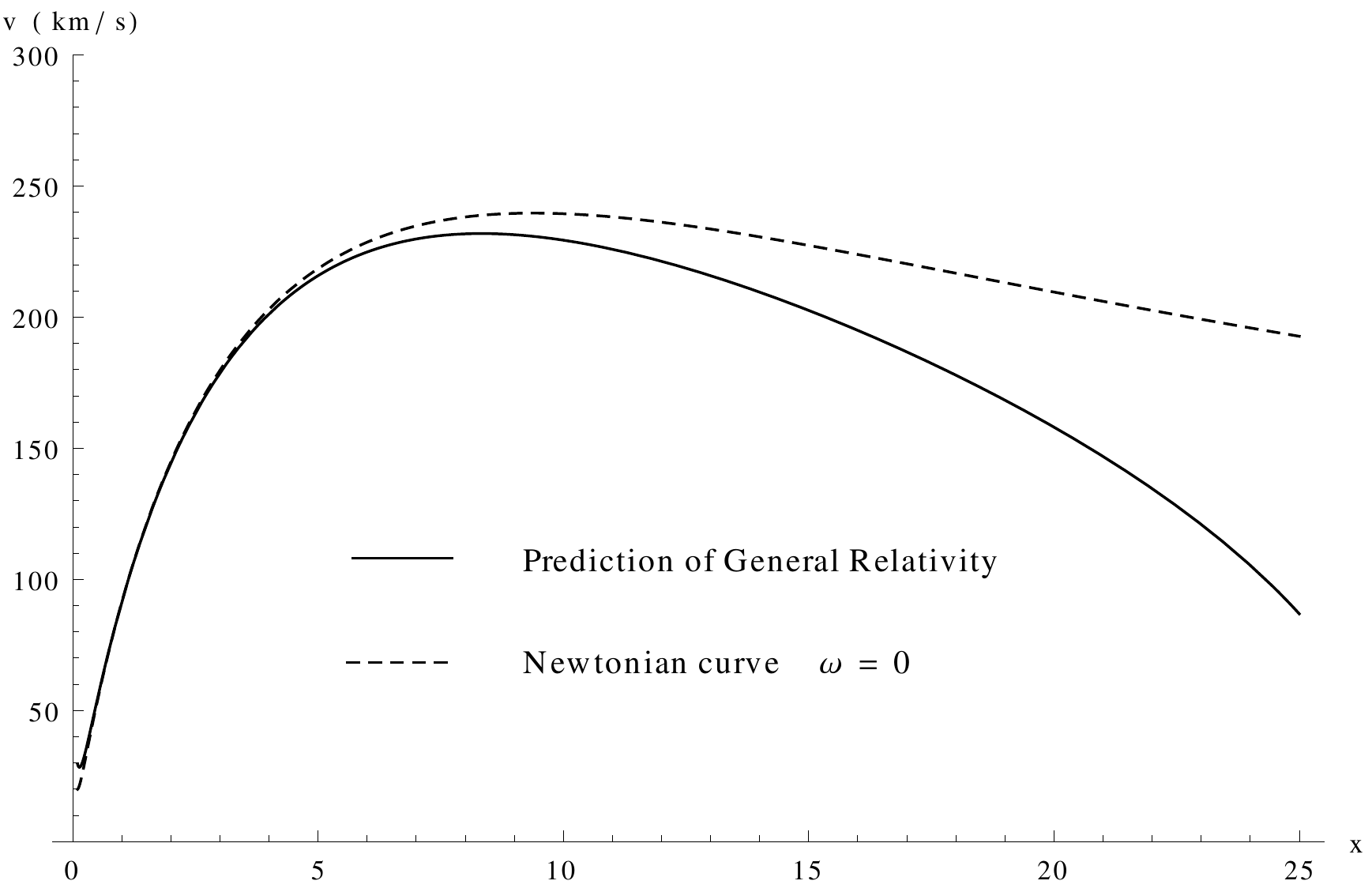}
\captionof{figure}{Faster than Newtonian curve}

\includegraphics[width=220 pt]{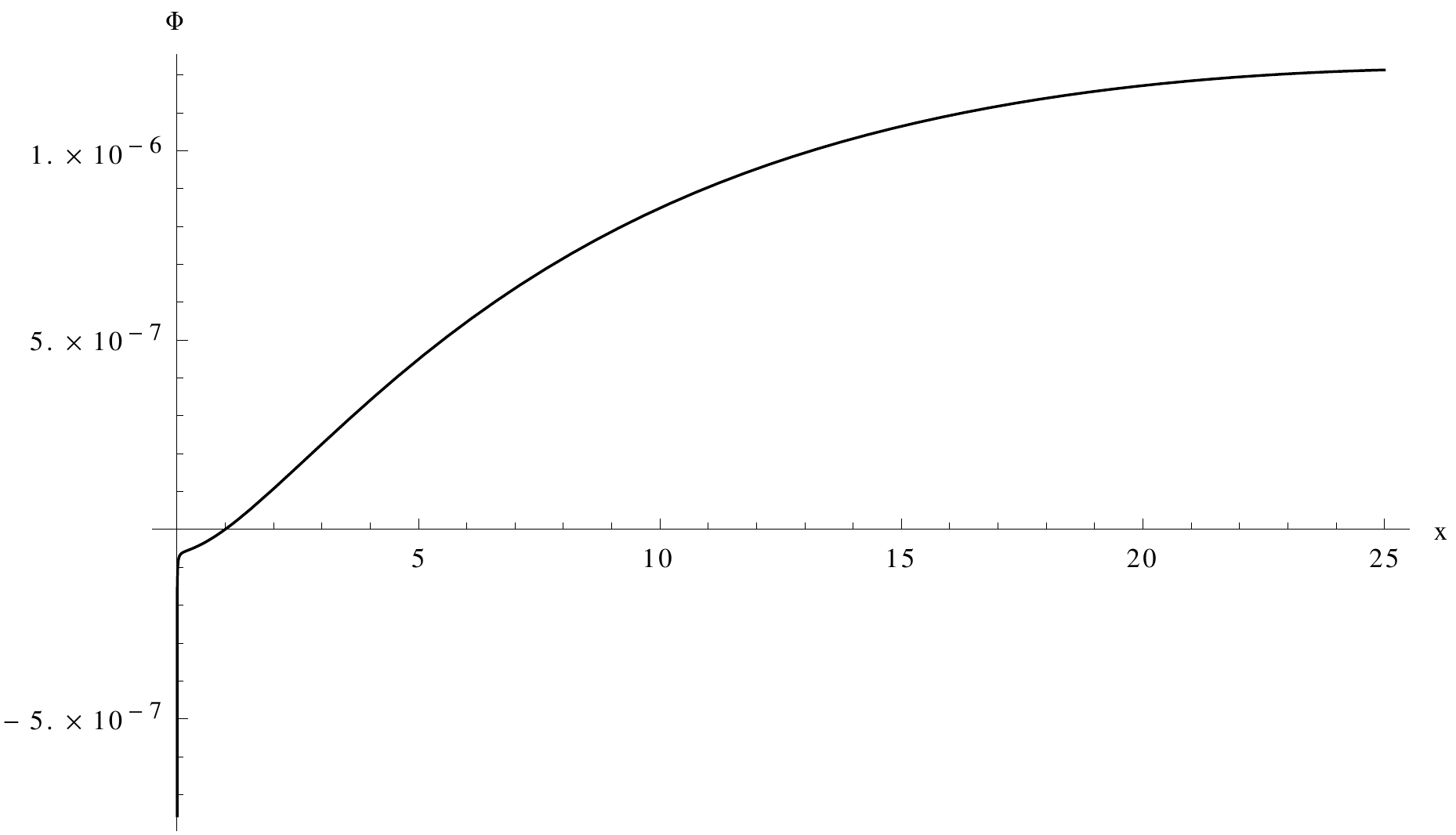}
\captionof{figure}{Faster than Newtonian potential}

\nopagebreak{
\includegraphics[width=200 pt]{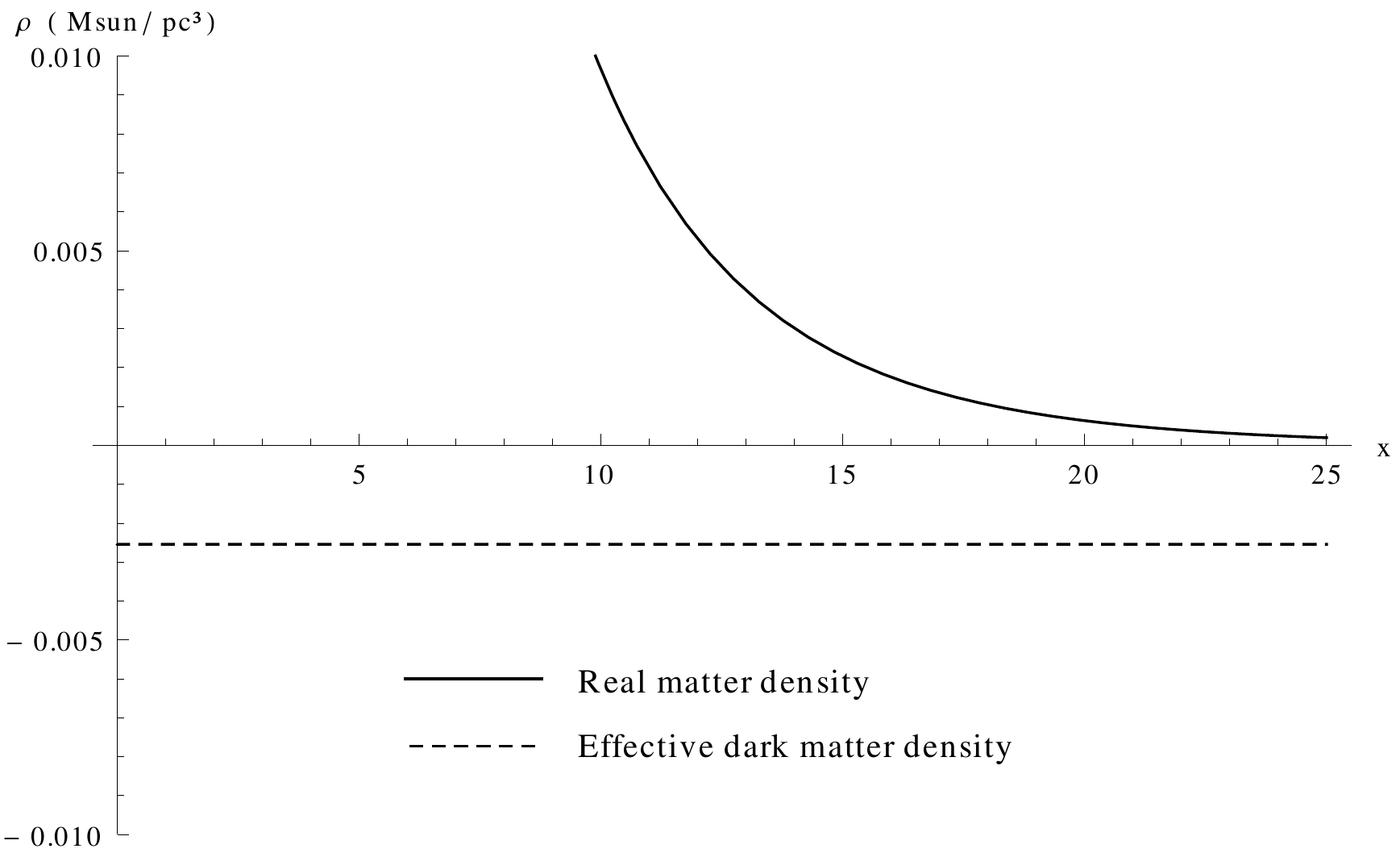}
\captionof{figure}{Matter density with effective anti-gravitational contribution}

\captionof{table}{Curve parameters}
\begin{tabular}{ l l | r l }
  $\omega$ & $-2\cdot10^{-50}$ Hz & $w(1)$ & $29$ m \\
  $r_0$ & $1$ kpc & $\mathit{v}(1)$ & $91$ $kms^{-1}$ \\ \cline{3-4}
  $\rho_0$ & $1$ $M_\odot pc^{-3}$ & Total galactic mass\\
  $n$ & $3/2$ &  $2\cdot 10^{11}$ $M_\odot$  \\
\end{tabular}
}

\end{center}

\section{Fitting real rotation curves}
Up to this point it has been shown how equation \eqref{key} is able to explain general behaviours on a qualitative ground. It is indeed positively surprising how this model can fit pretty well real rotation curves even very near to the galactic centre where simple S\'ersic profiles are expected to be less accurate because of the presence of the central galactic bulge. The fits hereby presented are a first study of a few well documented rotation curves. 

The fitting procedure is quite delicate because of the great number of parameters involved and no assurance is given on whether the set of best fitting parameters be unique. It is however very important that this model seems to dispose of virtually all major fitting difficulties appearing when using usual Newtonian gravitation. The best guide to the choice of starting parameters is the small local minimum region that many rotation curves possess near the origin, for the wideness of this region is very sensitive to $r_0$ and $\mathit{v}(1)$. 

This is a preliminary study and its main aim is to show the power of equation \eqref{key} when it comes to reproduce experimental behaviour. Further studies should focus on an independent determination of observational parameters such as $r_0$ and $n$ from luminosity profiles in order to eliminate ambiguities arising in the choice of starting fit parameters.
Hereby the theoretical and experimental rotation curve is given for each galaxy considered, as well as the Newtonian prediction accompanied by the plot of the matter distribution used for the fit and its effective dark matter counterpart. All parameters used for the fit are reported and for each galaxy an order of magnitude for the total galactic mass is given using the rough approximation that the galaxy be spherically symmetric.

\paragraph*{Cases with positive effective dark matter density}

\begin{center}
\captionof{figure}{NGC3198 Experimental data represent spectral measurements tracing CO lines for precise core measurements as well as HI and optical band, see \citep{Sofue} for reference. Note,  the size of experimental data points is not related to experimental uncertainty.}
\includegraphics[width=200 pt]{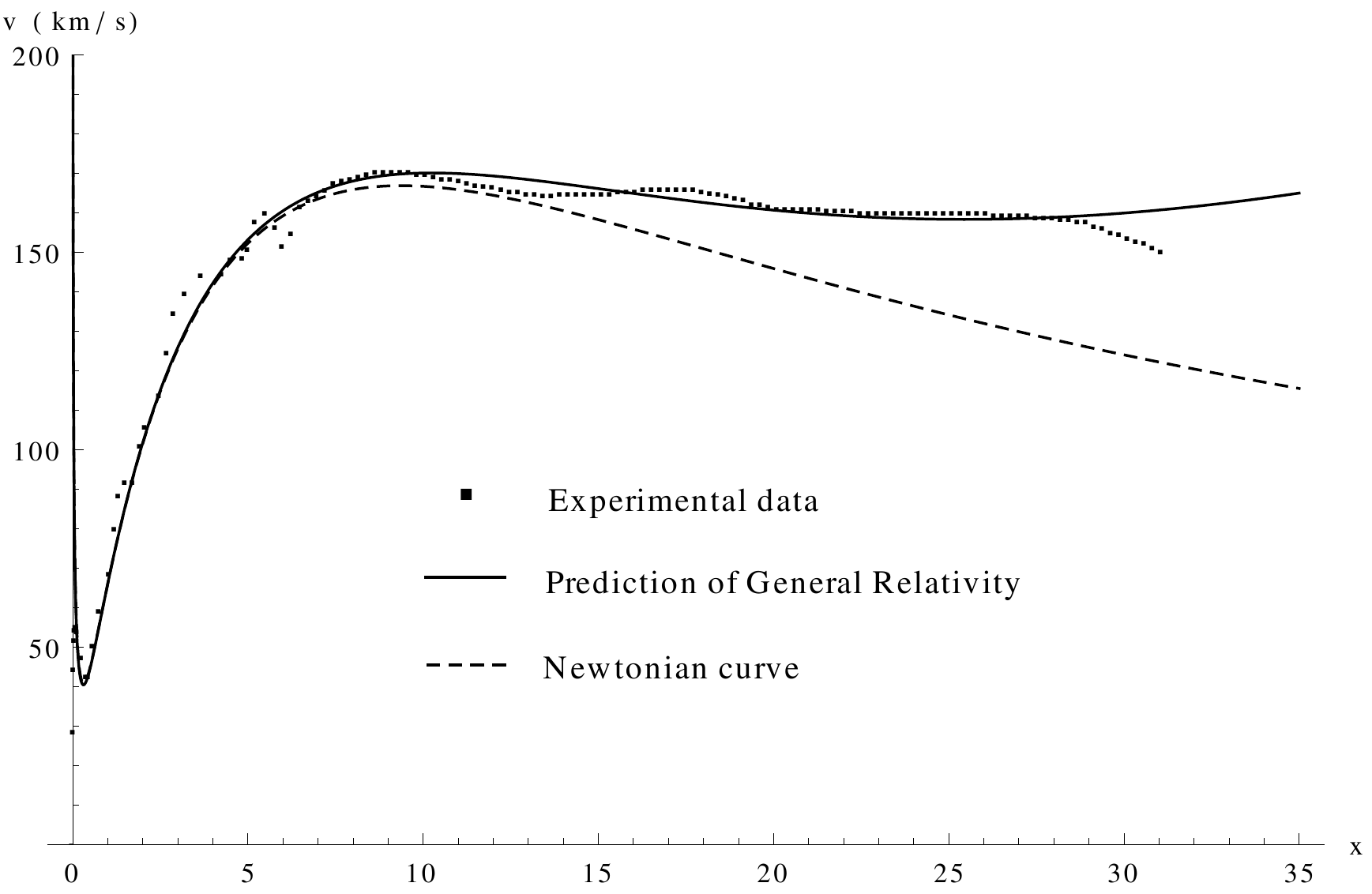}

\includegraphics[width=200 pt]{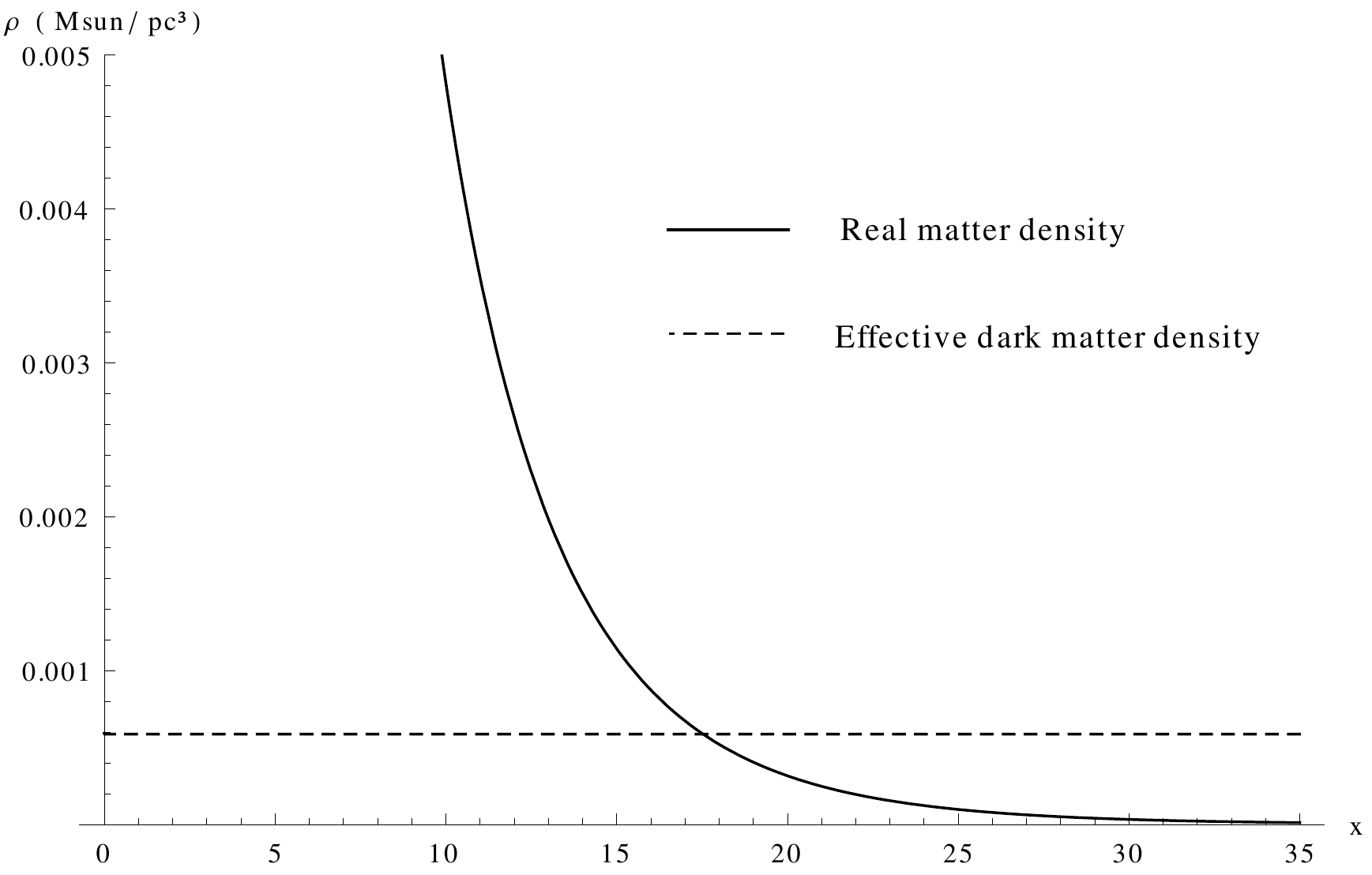}
\captionof{table}{Fitting parameters for NGC3198} 
\begin{tabular}{ l l | r l }
  $\omega$ & $4.2\cdot10^{-51}$ Hz & $w(1)$ & $28$ m \\
  $r_0$ & $1.0$ kpc & $\mathit{v}(1)$ & $66$ $kms^{-1}$ \\ \cline{3-4}
  $\rho_0$ & $0.5$ $M_\odot pc^{-3}$ & Galactic mass\\
  $n$ & $1.5$ &  $1.1\cdot 10^{11}$ $M_\odot$  \\
\end{tabular}
\begin{equation*}
\phantom{\int}
\end{equation*}
\captionof{figure}{NGC3521 Experimental data represent spectral measurements tracing CO lines for precise core measurements as well as HI and optical band, see \citep{Sofue} for reference.}
\includegraphics[width=200 pt]{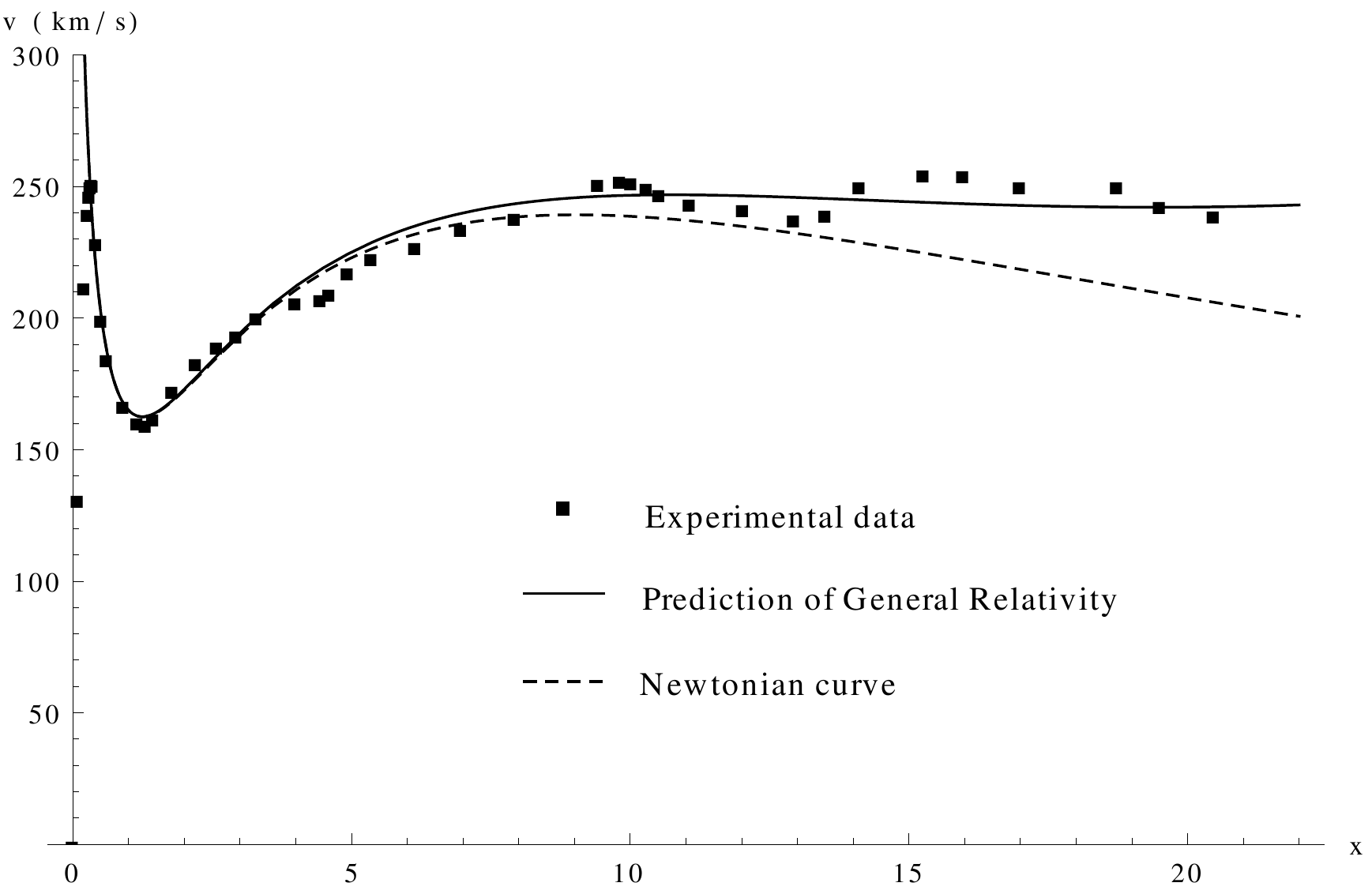}

\includegraphics[width=200 pt]{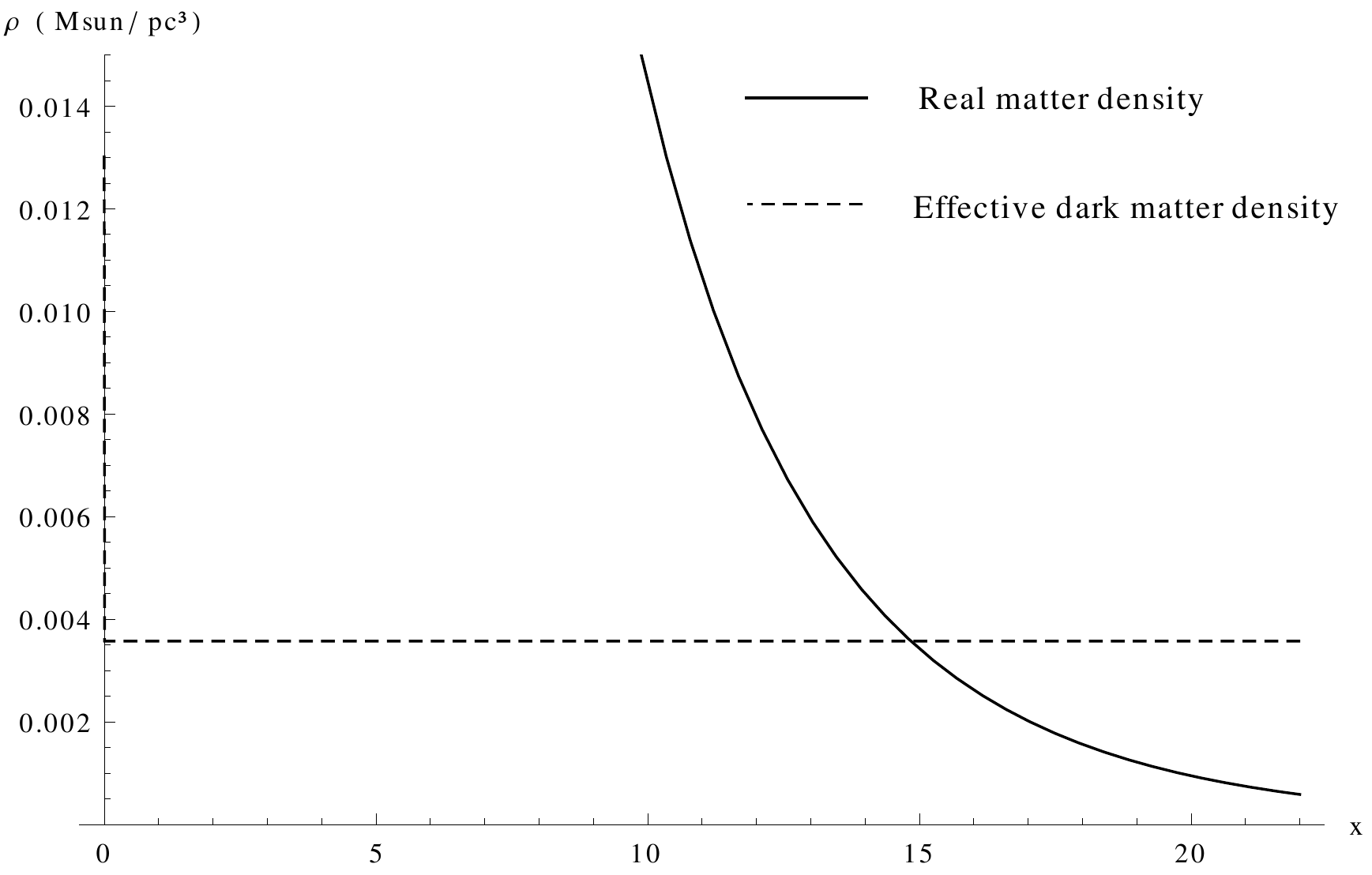} 

\captionof{table}{Fitting parameters for NGC3521} 
\begin{tabular}{ l l | r l }
  $\omega$ & $4.0\cdot10^{-50}$ Hz & $w(1)$ & $29$ m \\
  $r_0$ & $0.8$ kpc & $\mathit{v}(1)$ & $165$ $kms^{-1}$ \\ \cline{3-4}
  $\rho_0$ & $1.5$ $M_\odot pc^{-3}$ & Galactic mass\\
  $n$ & $1.5$ &  $1.7\cdot 10^{11}$ $M_\odot$  \\
\end{tabular}

\captionof{figure}{NGC5033 Experimental data represent spectral measurements tracing CO lines for precise core measurements as well as HI and optical band, see \citep{Sofue} for reference.}
\includegraphics[width=200 pt]{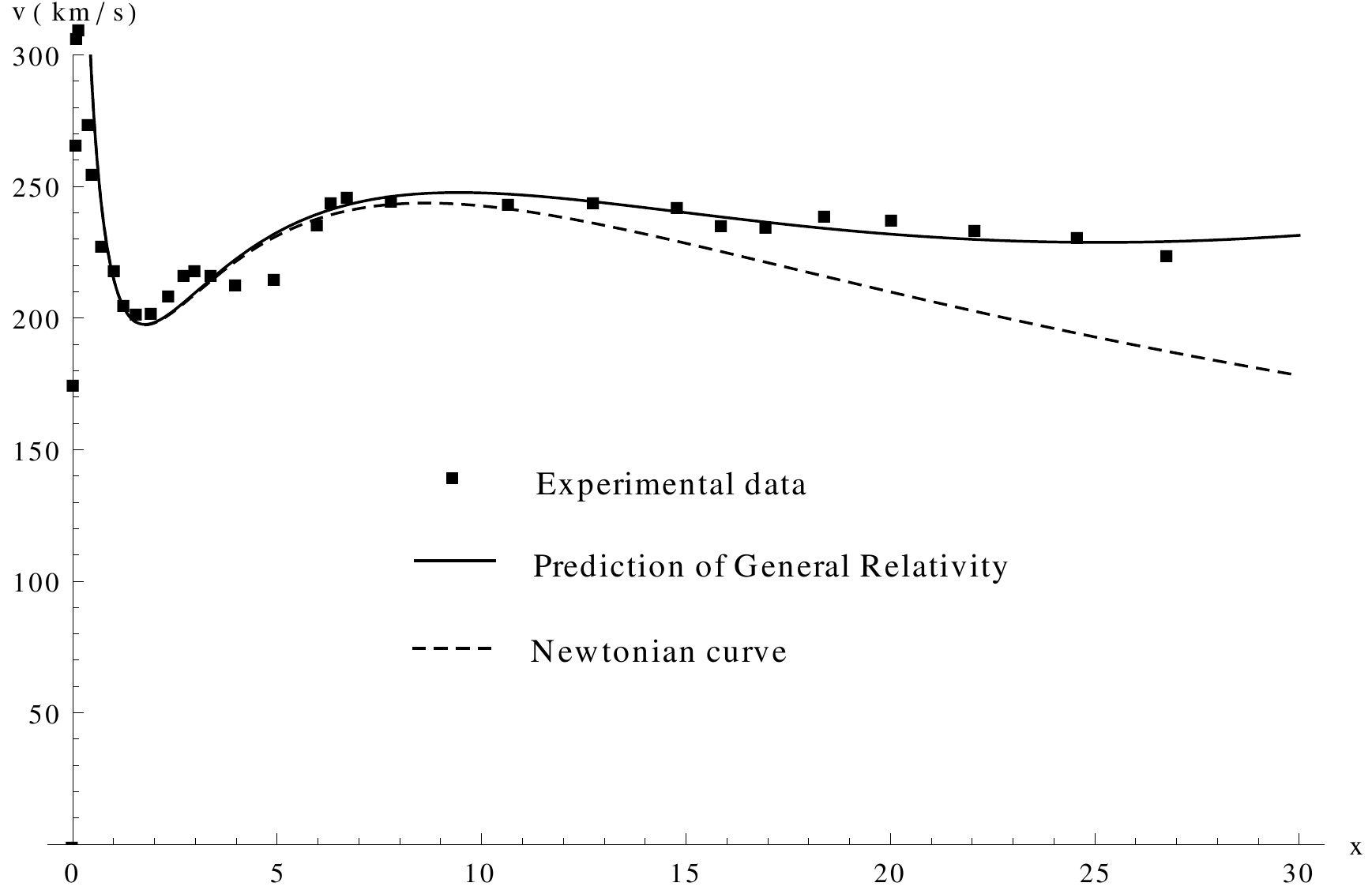}

\includegraphics[width=200 pt]{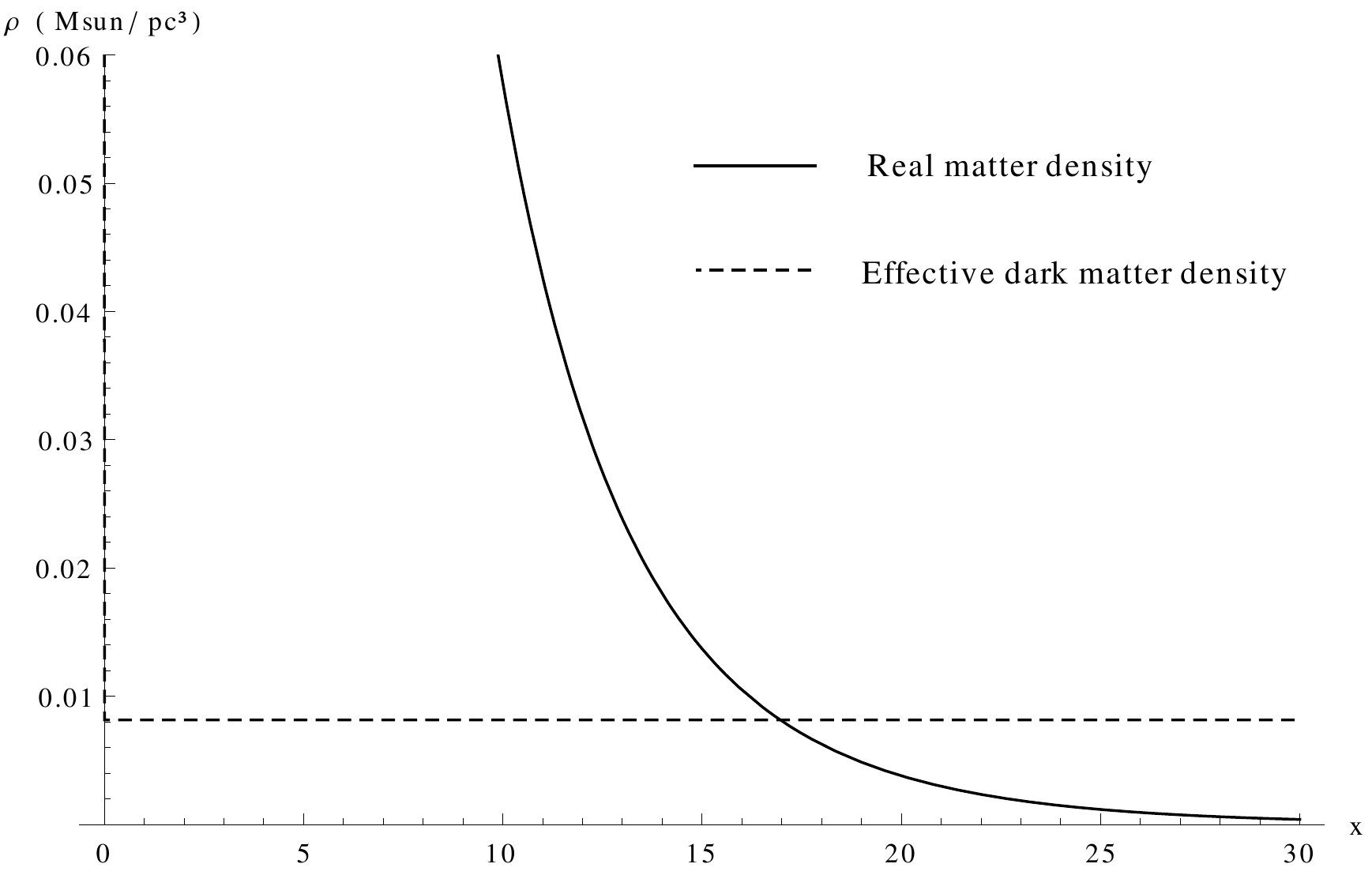}
\pagebreak

\captionof{table}{Fitting parameters for NGC5033} 
\begin{tabular}{ l l | r l }
  $\omega$ & $4.0\cdot10^{-49}$ Hz & $w(1)$ & $29$ m \\
  $r_0$ & $0.4$ kpc & $\mathit{v}(1)$ & $215$ $kms^{-1}$ \\ \cline{3-4}
  $\rho_0$ & $6.0$ $M_\odot pc^{-3}$ & Galactic mass\\
  $n$ & $1.5$ &  $8.4\cdot 10^{10}$ $M_\odot$  \\
\end{tabular}
\vfill
\captionof{figure}{NGC3495 Experimental data represent spectral measurements tracing CO lines for precise core measurements as well as HI and optical band, see \citep{Sofue} for reference.}
\includegraphics[width=200 pt]{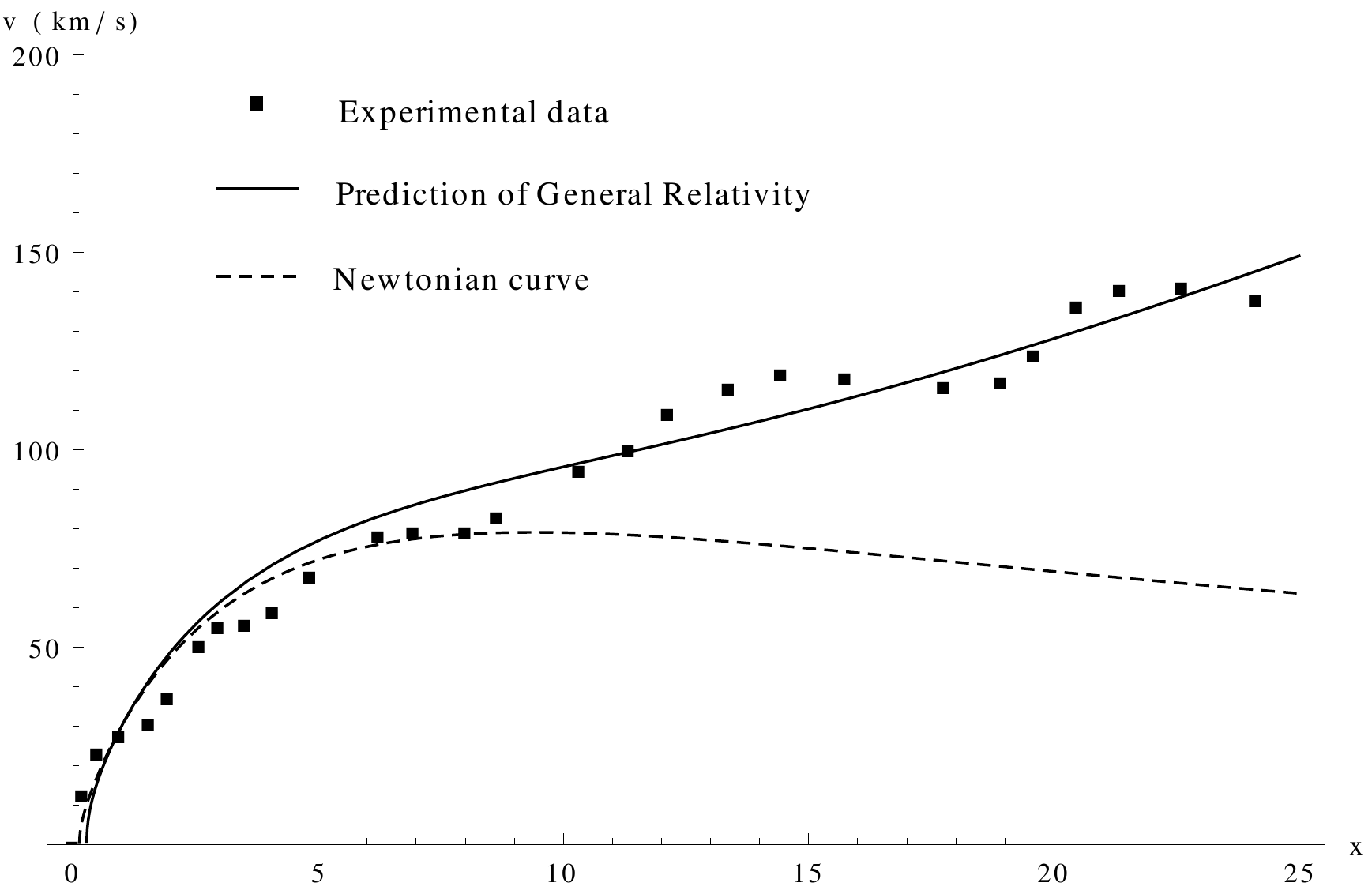}

\includegraphics[width=200 pt]{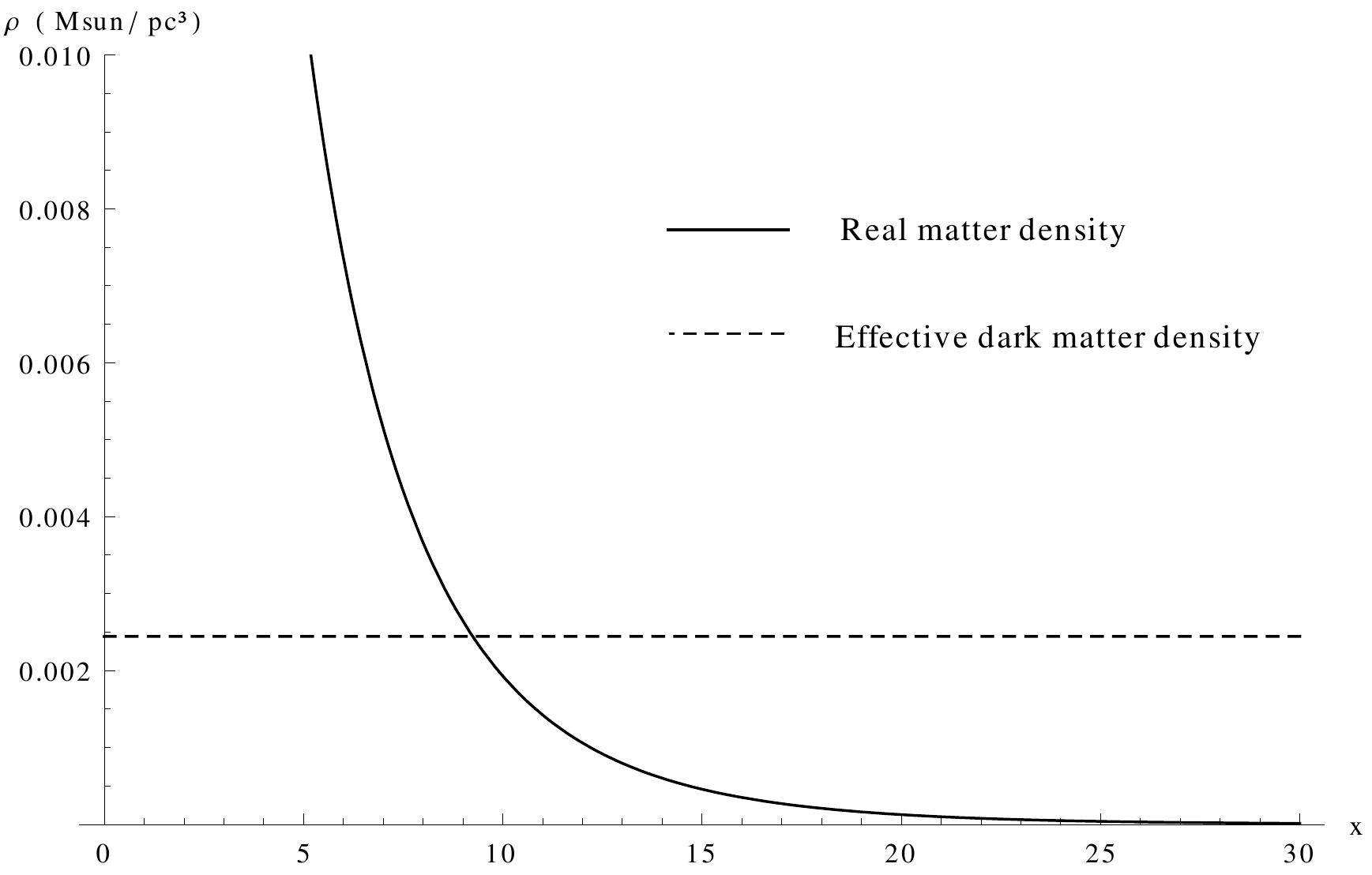}

\captionof{table}{Fitting parameters for NGC349} 
\begin{tabular}{ l l | r l }
  $\omega$ & $3.0\cdot10^{-50}$ Hz & $w(1)$ & $29$ m \\
  $r_0$ & $0.8$ kpc & $\mathit{v}(1)$ & $33$ $kms^{-1}$ \\ \cline{3-4}
  $\rho_0$ & $0.2$ $M_\odot pc^{-3}$ & Galactic mass\\
  $n$ & $1.5$ &  $2.2\cdot 10^{10}$ $M_\odot$  \\
\end{tabular}

\end{center}

\paragraph*{Cases with negative effective dark matter density}
The two following cases are examples found after a quick search in the literature, no decisive clue concerning the origin of their unusual behaviour has as yet been found. These are thus to be taken as an illustration regarding the possibility to fit such curves even when no significant perturbation coming from nearby objects or internal dynamics seems to have caused this anomalous behaviour.

\pagebreak
\begin{center}

\captionof{figure}{NGC864 Experimental data represent spectral measurements of the HI line, see \citep{Espada} for reference.}
\includegraphics[width=200 pt]{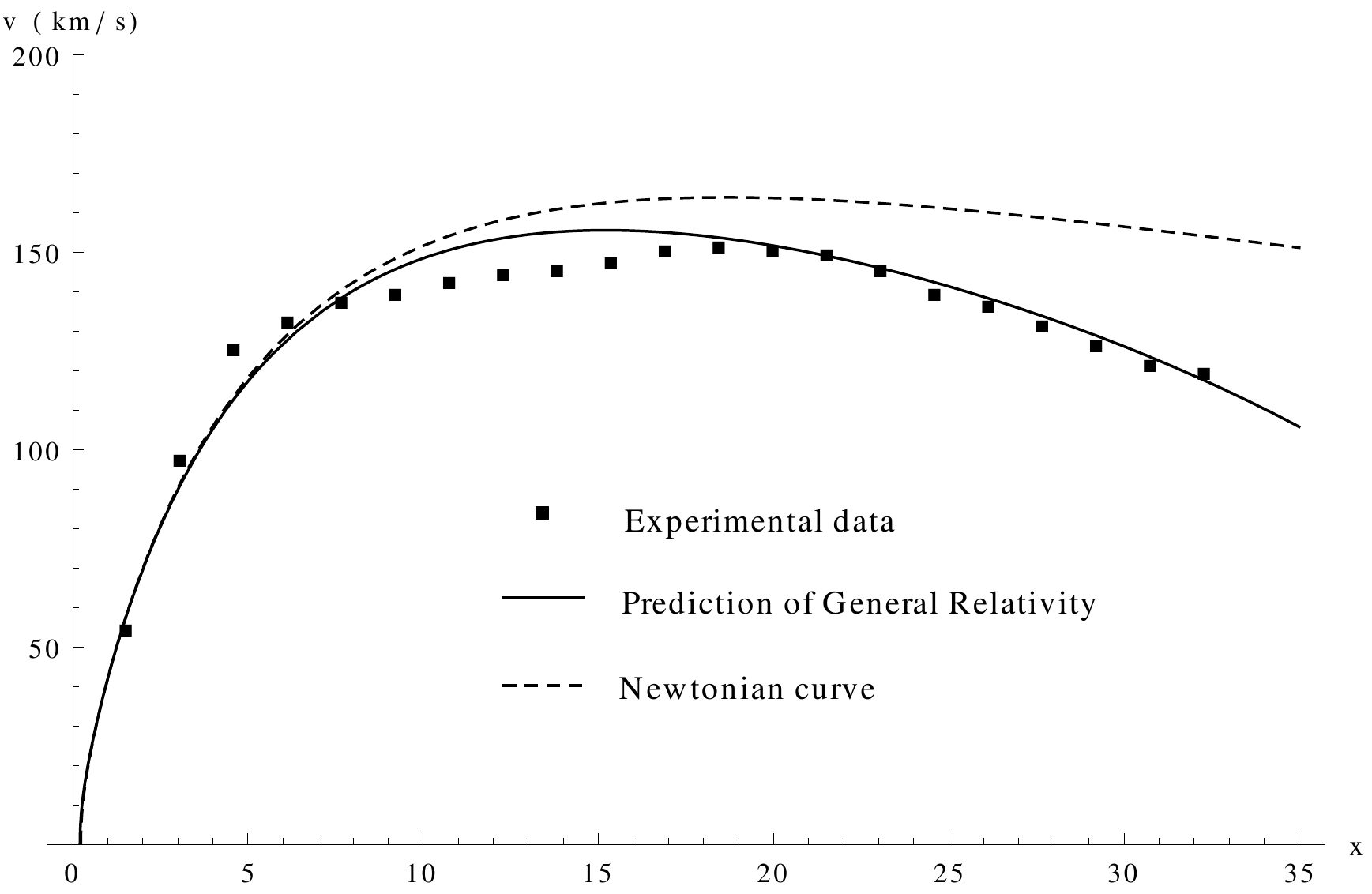}

\includegraphics[width=200 pt]{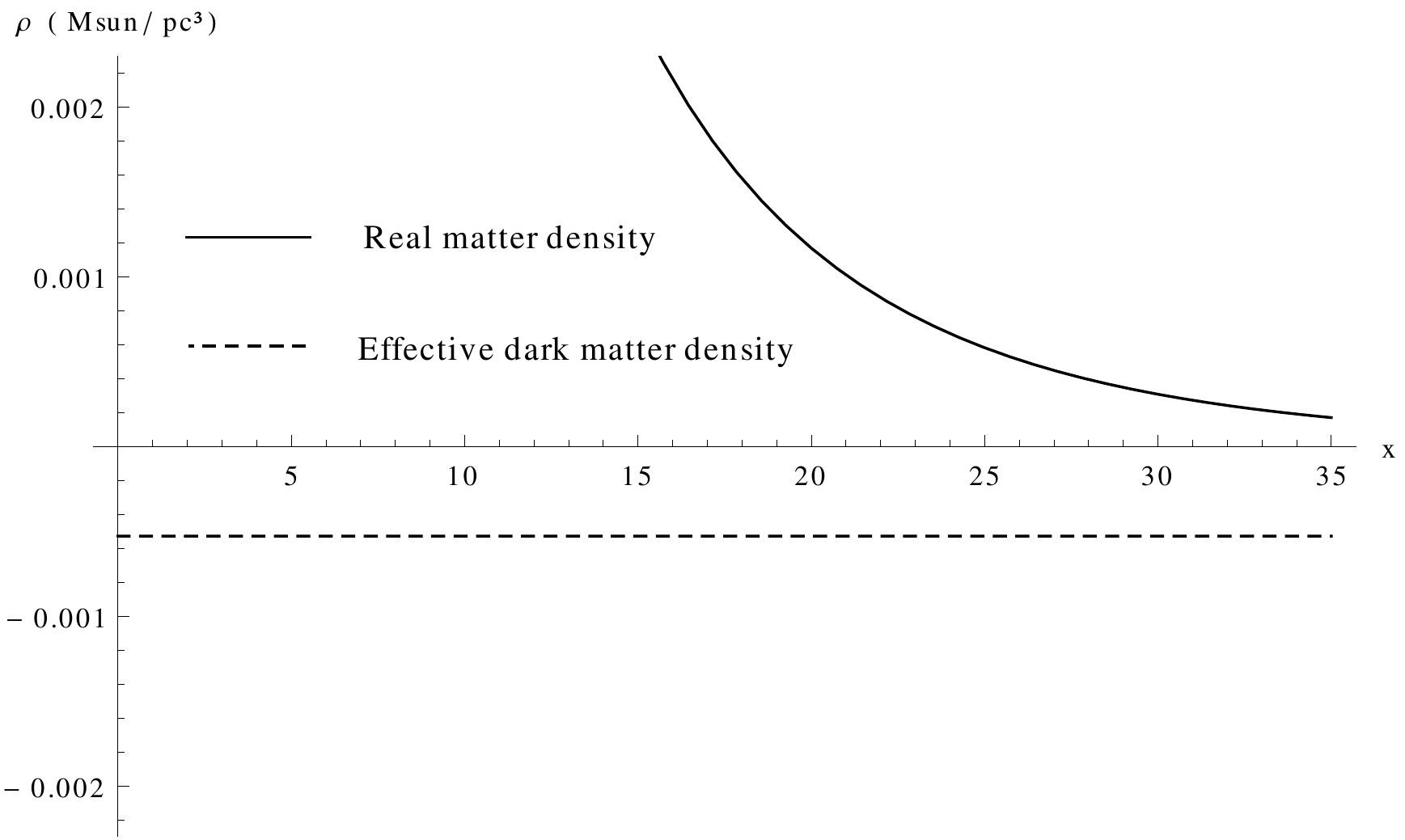} \quad

\captionof{table}{Fitting parameters for NGC864}
\begin{tabular}{ l l | r l }
  $\omega$ & $-4.5\cdot10^{-50}$ Hz & $w(1)$ & $100$ m \\
  $r_0$ & $1.0$ kpc & $\mathit{v}(1)$ & $42$ $kms^{-1}$ \\ \cline{3-4}
  $\rho_0$ & $0.23$ $M_\odot pc^{-3}$ & Galactic mass\\
  $n$ & $1.8$ &  $2.3\cdot 10^{11}$ $M_\odot$  \\
\end{tabular}

\captionof{figure}{AGC400848 Experimental data represent spectral measurements of the H$\alpha$ line, see \citep{Catinella} for reference.}

\includegraphics[width=200 pt]{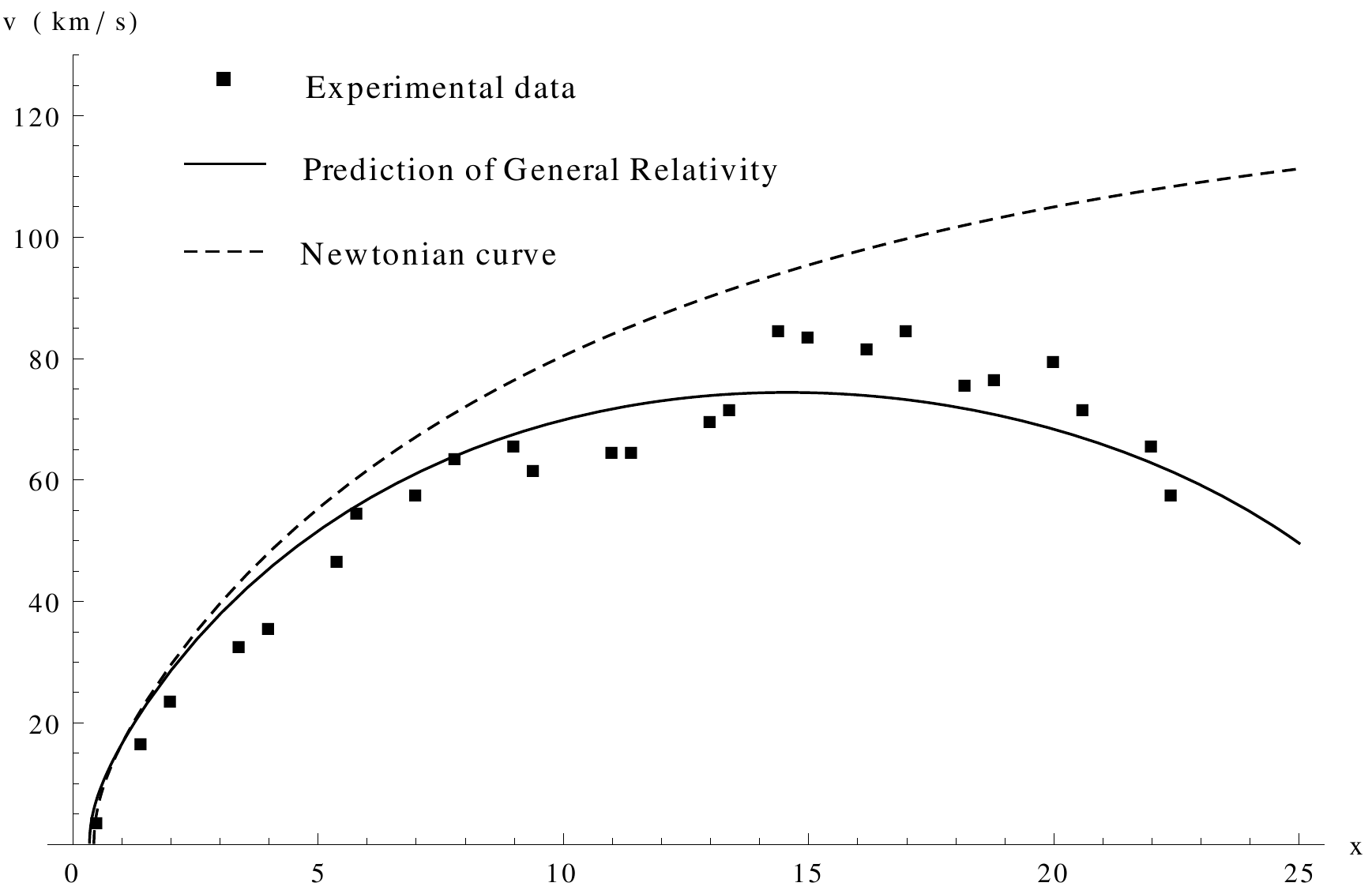}

\includegraphics[width=200 pt]{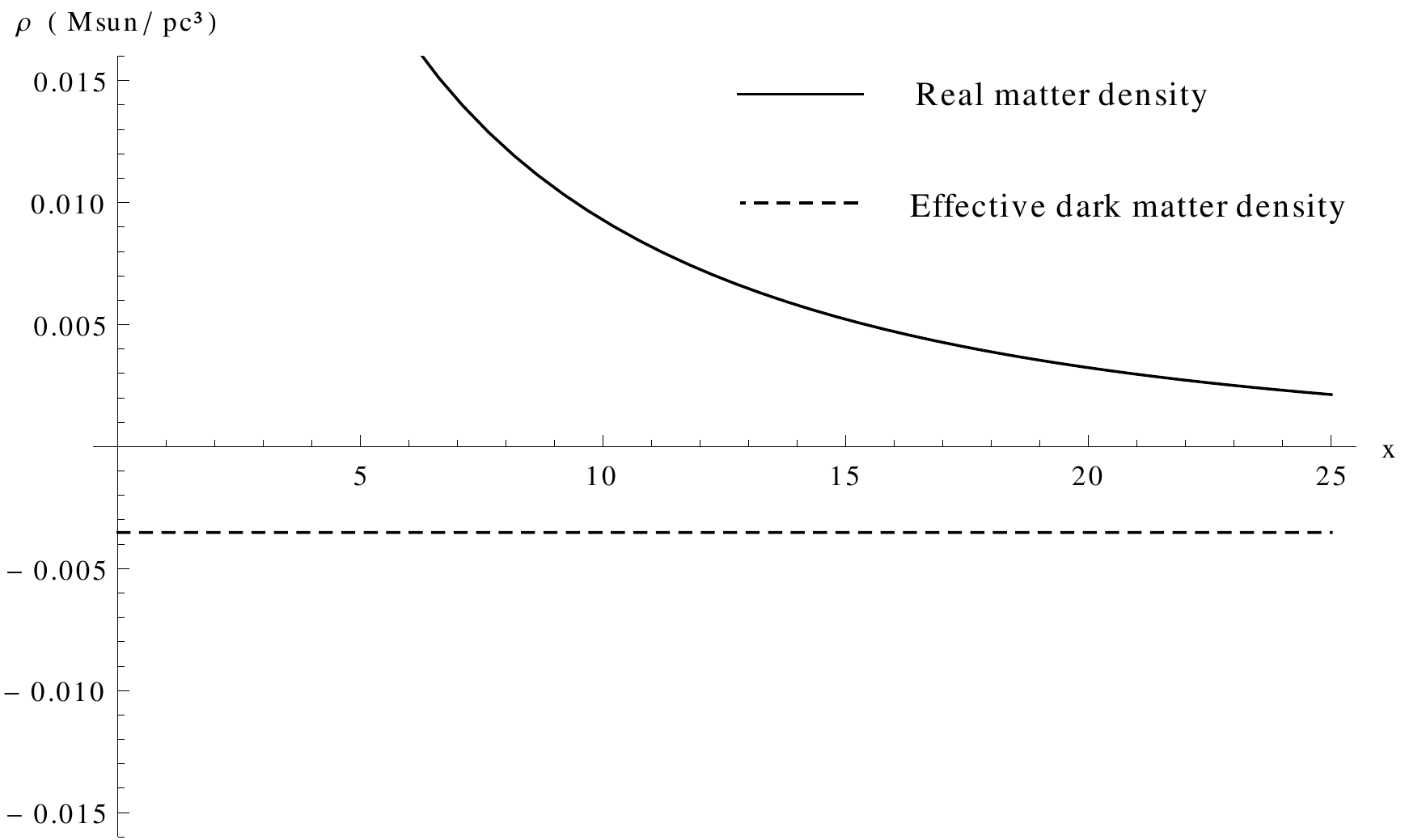}

\pagebreak
\captionof{table}{Fitting parameters for AGC400848}
\begin{tabular}{ l l | r l }
  $\omega$ & $-1.2\cdot10^{-48}$ Hz & $w(1)$ & $100$ m \\
  $r_0$ & $0.5$ kpc & $\mathit{v}(1)$ & $16.5$ $kms^{-1}$ \\ \cline{3-4}
  $\rho_0$ & $0.16$ $M_\odot pc^{-3}$ & Galactic mass\\
  $n$ & $2.2$ &  $1.9\cdot 10^{11}$ $M_\odot$  \\
\end{tabular}

\end{center}

\section{Conclusions}
In this work I have shown how General Relativity is perfectly able to explain galactic rotation curves without any need for dark matter. The latter is seen as a purely dynamical consequence of the expansion of our Universe; it seems to be caused by a non-linear coupling between the background expanding Roberson-Walker metric and the nearly Newtonian cosmological perturbation placed in it. This is important because, if the structure of some galaxy were sufficiently well known, one could, as a matter of principle, measure Hubble's constant studying its internal dynamics. Like Foucault measured the Earth's rotation standing in a Parisian drawing room, one could measure the cosmological expansion while pleasantly staying in our solar system, solely watching its spin around the centre of the Milky Way.

In this work I have also stressed the point of how the non-linearities of Einstein's equations are crucial when it comes to understand many interesting features of gravitation, have clearly shown how these act in the nearly Newtonian limit through the appearance of effective dark matter density terms and also shown how these get lost during linearisation.

The fact that a very simple and reasonable model, such as exponential matter distributions, produces good fits using this approach confirms the interest of this scheme of thought. This approach seems also able to explain to some extent the dominance of dark matter in low-mass galaxies and confirms the inverse dependence of the galaxies' dark matter content fraction on its luminosity. It also gives good hopes that more physically refined models using equation \eqref{key} might be very useful to interpret a wide range of phenomena as yet unexplained within standard General Relativity. Equation \eqref{key} opens the way to a reconsideration of many astrophysical phenomena. For example, as predicted rotation curves depart significantly from the usual Newtonian ones, one expects predictions regarding galactic formation and evolution to be significantly altered too. In order to proceed to such a study one would have to introduce a suitable equation of state for the galactic constituents. Of course Galactic clusters can be treated in exactly the same way. This approach is also expected to be important when studying gravitational lensing effects that also need the presence of galactic dark matter to explain observations, albeit in this case one is in need of a higher-order tractation because equation \eqref{key} leaves out all fully (special) relativistic phenomena.

\section*{Contextual Notes} This work has been born from the line of thought in which effects such as the flatness of rotation curves seek an explanation in terms of the gravitational action of faraway matter in an expanding universe. A representative work of this line is that of \citet{Carati}. They use a discrete matter distribution in a Minkowskian space-time and impose Hubble's law on it, physical results are then deduced using mean field methods, in contrast to the continuum approach used here based on Einstein's equations. In their work, they very interestingly obtain the spatially flat Robertson-Walker metric as mean metric ($\kappa=0$), using this fact as a consistency condition, and are able to fit galactic rotation curves. Among these are the two above mentioned cases with faster than Newtonian decrease.
A deeper mathematical connection between the present work and theirs has as yet to be worked upon. 

\begin{acknowledgements}
I am most grateful to professors Andrea Carati and Luigi Galgani who have aroused my interest in this field. I must thank them for the long, interesting and fruitful conversations we have had in their study and for their help and support during my work. I especially thank Prof. Carati whose interest in my previous thesis work has led to our interesting exchange, helped me greatly, and whose original ideas have fostered me not only on a technical ground but also in my approach to scientific thought. I also thank Dr T. M. Nieuwenhuizen for his constructive critique.
\end{acknowledgements}


\bibliographystyle{aa} 
\bibliography{Bibliography} 

\begin{thebibliography}{15}
\expandafter\ifx\csname natexlab\endcsname\relax\def\natexlab#1{#1}\fi

\bibitem[{Beckenstein(2004)}]{Beck}
Beckenstein, J.~D. 2004, Physical Revew D, 70, 083509

\bibitem[{Bertin \& van Albada(1998)}]{Bert}
Bertin, G. \& van Albada, J.~S. 1998, {E}nciclopedia del {N}ovecento {I}{I}
  {S}upplemento (Istituto Giovanni Treccani)

\bibitem[{Burkert(1995)}]{Burk}
Burkert, A. 1995, The Astrophysical Journal, 447, L25

\bibitem[{Caon {et~al.}(1993)Caon, Capaccioli, \& D'Onofrio}]{Sers}
Caon, N., Capaccioli, M., \& D'Onofrio, M. 1993, Monthly Notices of the Royal
  Astronomical Society, 265, 1013

\bibitem[{Carati \& Galgani(2011)}]{Carati}
Carati, A. \& Galgani, L. 2011, Asociaci\`on Argentina de Astronom\`ia

\bibitem[{Catinella {et~al.}(2005)Catinella, Haynes, \& Giovanelli}]{Catinella}
Catinella, B., Haynes, M., \& Giovanelli, R. 2005, Astrononomical Journal, 452,
  455

\bibitem[{Espada {et~al.}(2005)Espada, Bosma, L., Athanassoula, Leon, Sulentic,
  \& Yun}]{Espada}
Espada, D., Bosma, A., L., V.-M., {et~al.} 2005, Astronomy \& Astrophysics,
  452, 455

\bibitem[{Jardel \& Gebhardt(2012)}]{Fornax}
Jardel, J.~R. \& Gebhardt, K. 2012, The Astrophysical Journal, 746, 89

\bibitem[{K. \& Currie(1994)}]{bisSers}
K., Y.~C. \& Currie, M.~J. 1994, Monthly Notices of the Royal Astronomical
  Society, 268, L11

\bibitem[{Massey {et~al.}(2010)Massey, Kitchny, \& Richard}]{Lense}
Massey, R., Kitchny, T., \& Richard, J. 2010, Reports on Progress in Physics,
  73, 886901

\bibitem[{Milgrom(1983)}]{Milgr}
Milgrom, M. 1983, The Astrophysical Journal, 270, 365

\bibitem[{Miscellaneous(1970)}]{Ein}
Miscellaneous. 1970, {A}lbert {E}instein: {P}hilosopher--{S}cientist, 2nd edn.,
  Vol.~3 (Northwestern University \& Southern Illinois University, USA: Paul
  Arthur Schilpp)

\bibitem[{Navarro {et~al.}(1996)Navarro, Frenk, \& White}]{CDM}
Navarro, J.~F., Frenk, C.~S., \& White, S. D.~M. 1996, The Astrophysical
  Journal, 462, 563

\bibitem[{Persic {et~al.}(1996)Persic, Salucci, \& Stel}]{Luminosity}
Persic, M., Salucci, P., \& Stel, F. 1996, Monthly Notices of the Royal
  Astronomical Society, 281, 26

\bibitem[{Sofue(1997)}]{Sofue}
Sofue, Y. 1997, Publications of the Astronomical Society of Japan, 49

\end{thebibliography}

\end{document}